\documentclass[pteplogo]{ptephy_v1}%%%%%% to generate preprint number with ptep logo
\preprintnumber{1803-010} %%% %%% Insert preprint number here

\usepackage{lineno}

\usepackage{rotating}
\usepackage{textcomp}
\usepackage{amsmath}
\usepackage[justification=centering]{subcaption}

\begin{document}
 
\title{Development of new radon monitoring systems in the Kamioka mine}

%%%%%%%%%%%%%%%%%%%%%%%%%%%%%%%%%%%%%%%%%%%%%%%%%%%%%%%%%%
%%			Author list			%%
%%%%%%%%%%%%%%%%%%%%%%%%%%%%%%%%%%%%%%%%%%%%%%%%%%%%%%%%%%

\author[1]{G.~Pronost\thanks{E-mail: pronost@km.icrr.u-tokyo.ac.jp}}

\author[1]{M.~Ikeda}

\author[2]{T.~Nakamura}

\author[1,3]{H.~Sekiya}

\author[1]{S.~Tasaka}

%%%%%%%%%%%%%%%%%%%%%%%%%%%%%%%%%%%%%%%%%%%%%%%%%%%%%%%%%%
%%			Institutions list		%%
%%%%%%%%%%%%%%%%%%%%%%%%%%%%%%%%%%%%%%%%%%%%%%%%%%%%%%%%%%

\affil[1]{Kamioka Observatory, Institute for Cosmic Ray Research, University of Tokyo, Kamioka, Gifu 506-1205, Japan}
\affil[2]{Department of Physics, Gifu University, Gifu, Gifu 501-1193, Japan}
\affil[3]{Kavli Institute for the Physics and
Mathematics of the Universe (WPI), The University of Tokyo Institutes for Advanced Study,
University of Tokyo, Kashiwa, Chiba 277-8583, Japan }

%%%%%%%%%%%%%%%%%%%%%%%%%%%%%%%%%%%%%%%%%%%%%%%%%%%%%%%%%%
%%			Abstract			%%
%%%%%%%%%%%%%%%%%%%%%%%%%%%%%%%%%%%%%%%%%%%%%%%%%%%%%%%%%%

\begin{abstract}
Radioactivity from radon is a major threat for high-precision low energy physics experiments like the ones in the Kamioka Mine.
We developed a new high sensitive radon monitoring system and conducted systematic radon concentration
measurements for the first time in Kamioka. The system consists of portable radon detectors
with a capacity of 1 L and new electronics based on Raspberry Pi \cite{Raspberry}. These radon detectors measure 
the radon in air with electro-static collection and a PIN photodiode. We measured the absolute humidity
dependence of the 1-L radon detector for air to be $C_{F} (A_{H}) = (12.86 \pm 0.40) - (1.66 \pm 0.19) \sqrt{A_{H}}$ $\mathrm{(counts/day)/(Bq/m^3)}$.
The background level of the 1-L radon detector is $0.65\pm0.15$ (stat.) count/day. This corresponds to a detection limit of $\sim0.4$ Bq/m$^3$ in 
a one-day measurement. Data was collected for a period of more than one year with twenty one 1-L radon detectors in the Kamioka mine. 
They indicate seasonal and day-night variations in radon concentration within the mine.
These results also allow us to confirm the stability of the new Raspberry Pi electronics. 
 
\end{abstract}

\subjectindex{xxxx, xxx}

\maketitle

%%%%%%%%%%%%%%%%%%%%%%%%%%%%%%%%%%%%%%%%%%%%%%%%%%%%%%%%%%
%%			Keywords			%%
%%%%%%%%%%%%%%%%%%%%%%%%%%%%%%%%%%%%%%%%%%%%%%%%%%%%%%%%%%

%\keywords{Dosimetry concepts and apparatus; Gaseous detectors; Radiation monitoring; Front-end electronics for detector readout}

%%%%%%%%%%%%%%%%%%%%%%%%%%%%%%%%%%%%%%%%%%%%%%%%%%%%%%%%%%
%%			Article 			%%
%%%%%%%%%%%%%%%%%%%%%%%%%%%%%%%%%%%%%%%%%%%%%%%%%%%%%%%%%%

\section{Introduction}

$^{222}$Rn (called radon hereafter) is a radioactive gas abundant in most of underground places. It is produced in the 4n+2 decay chain of $^{238}$U, so-called ``radium series''. 
Its decay chain is as follows: 
$^{222}_{\,86}\mathrm{Rn} \xrightarrow[3.82\,d]     {\alpha_1} \, 
 ^{218}_{\,84}\mathrm{Po} \xrightarrow[3.1\,min]    {\alpha_2} \, 
 ^{214}_{\,82}\mathrm{Pb} \xrightarrow[26.8\,min]   {\beta^-}  \, 
 ^{214}_{\,83}\mathrm{Bi} \xrightarrow[19.9\,min]   {\beta^-}  \, 
 ^{214}_{\,84}\mathrm{Po} \xrightarrow[164.3\,\mu s]{\alpha_3} \, 
 ^{210}_{\,82}\mathrm{Pb}$
                         $\xrightarrow[22.26 \, a]  {\beta^-}  \, 
 ^{210}_{\,83}\mathrm{Bi} \xrightarrow[5.01 \, d]   {\beta^-\ }\, 
 ^{210}_{\,84}\mathrm{Po} \xrightarrow[138.38\,d]   {\alpha_4} \,
 ^{206}_{\,82}\mathrm{Pb}$,
where $\alpha_{1}$, $\alpha_{2}$, $\alpha_{3}$, and $\alpha_{4}$ are the different $\alpha$ particles in this decay chain with energies of
$E(\alpha_1)=5.590$\,MeV, $E(\alpha_2)= 6.002$\,MeV, $E(\alpha_3)=7.833$\,MeV, and $E(\alpha_4)=5.407$\,MeV \cite{IsotopesRef}.
Due to the production of $\alpha$ and $\beta^-$ particles over a long period, $^{222}$Rn is a serious source of background for low-energy physics experiments and especially 
underground experiments. In addition, as a radioactive gas, workers in underground environments can be exposed to high concentration of radon, which can lead to serious 
health issues \cite{RadonRisk}. These two issues suggest the need to monitor the radon concentration in underground environments.

The Kamioka mine in the Gifu Prefecture of Japan hosts several high-precision, low-energy experiments, including Super-Kamiokande \cite{FreshAirPaper}, KamLAND \cite{KamLAND}, 
XMASS \cite{XMASS}, CANDLES \cite{CANDLES}, and NEWAGE \cite{NEWAGE}. To monitor the radon concentration in the air of the mine, we developed a 
precise and cost-effective system using a 1-L detector with Raspberry Pi-based electronics.

\section{Detector design}

To measure the radon concentration in the air, we used the electrostatic collection method of the daughter nuclei of $^{222}$Rn, combined 
with the energy measurement of alpha particles from the daughter-nuclei decays. This energy measurement is performed using a PIN  photo-diode. 
Previous publications\cite{Kotrappa} showed that $90\%$ of the isotopes produced in the $^{222}$Rn-nuclei-decay chain are positive ions. Therefore, they can 
be collected on a PIN photo-diode on which a negative High-Voltage (HV) is applied.

$\alpha_{2}$, $\alpha_{3}$,  and $\alpha_{4}$, referring to the $\alpha$ particles from the $^{218}$Po, $^{214}$Po, and $^{210}$Po decays, respectively, can be detected by the 
PIN photo-diode if the polonium nucleus has been collected. 
Since  $^{222}$Rn nuclei are not ions, they are unlikely to be collected on the PIN photo-diode, $\alpha_{1}$ are then unlikely to be detected. Only $\alpha_{2}$, 
$\alpha_{3}$,  and $\alpha_{4}$ measurements are performed.

\begin{figure}
	\centering
	\includegraphics[width=.6\paperwidth]{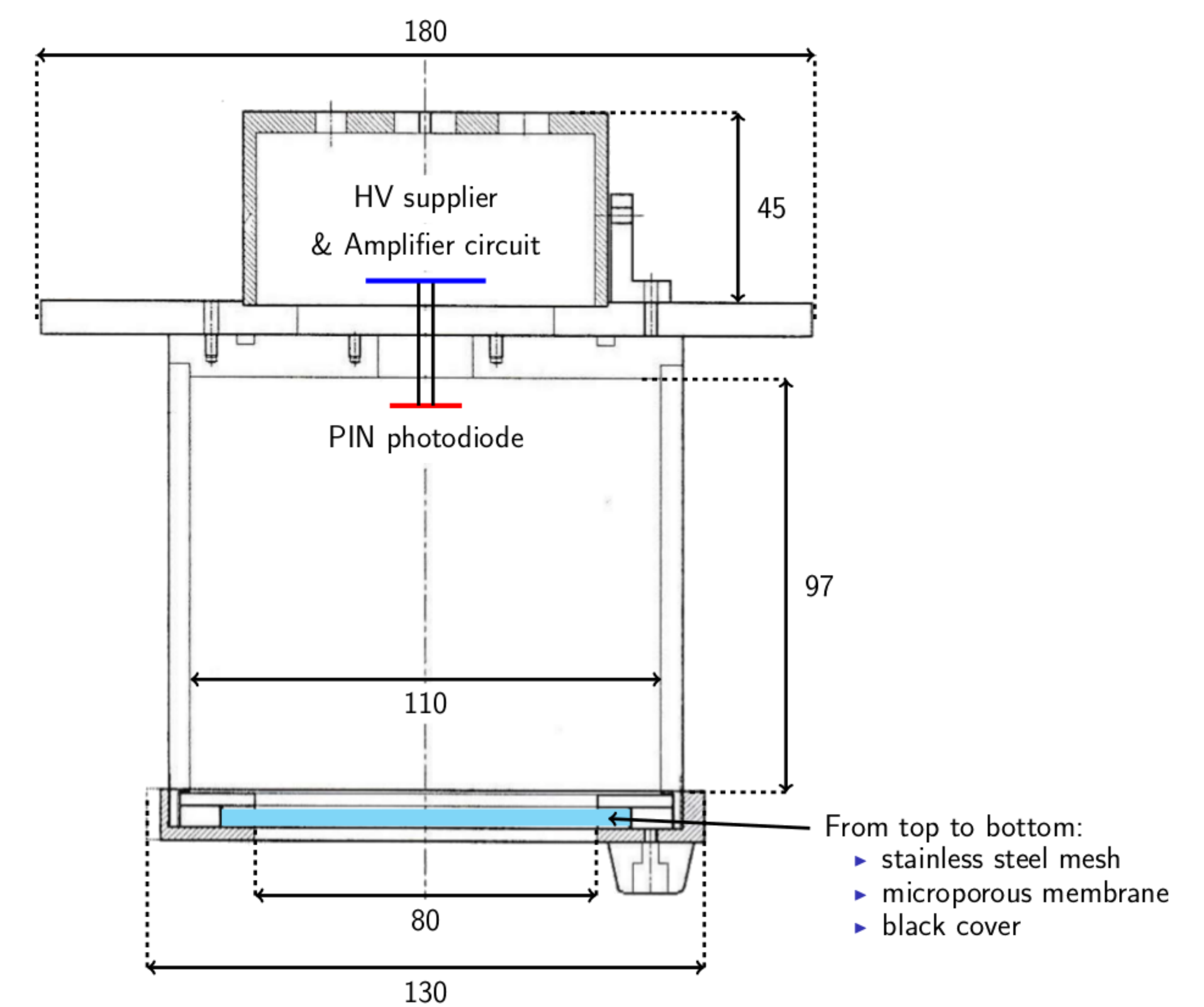}
	\caption{Schematic of a 1-L radon detector. Units are in mm.
	\label{fig:DesignRn}}
\end{figure}

Fig. \ref{fig:DesignRn} shows a schematic of the new 1-L radon detectors we use in the Kamioka Mine, this schematic is of the same type of that shown 
in Fig. 4 of \cite{DevSmall}. These detectors are made of stainless steel cases, opened at the 
bottom in order to allow air from the atmosphere to enter in the detector. A Hamamatsu S3590-09 PIN photo-diode is located at the top of the detector 
and  is connected via a feed-through to a HV supplier, a preamplifier and a shaping amplifier. 
 The air opening at the bottom of the detector is covered by three different layers. From inside to outside, these are a stainless-steel
mesh, a microporous membrane, and a black cover. The stainless-steel mesh is used as an electric ground in continuity with the detector case. It consists of 100 mesh per inch.
The microporous membrane is used as a filter to remove dust from the incoming air and thus the daughter nuclei of the radon. This allows us to 
collect only polonium, bismuth, and lead positive ions from radon decays occurring within the detector case. The microporous membrane used is a 
polypropylene membrane produced by CELGARD\textsuperscript{TM} company (product's name: CELGARD\textsuperscript{\textcopyright} 2400).
The last cover is a piece of black fabric, which is used to shield the inside of the detector, and thus the PIN photo-diode, from external light.

The ventilation rate of the 1-L radon detector has been calculated to be about seven times per hour with natural ventilation. The computation was done 
 using the theoretical equation and experimental values from \cite{Ventilation}. This large value is owing to the 80 mm diameter ventilation window.

\section{Electronics}

\begin{sidewaysfigure}
	\centering
	
	\includegraphics[width=.8\paperwidth]{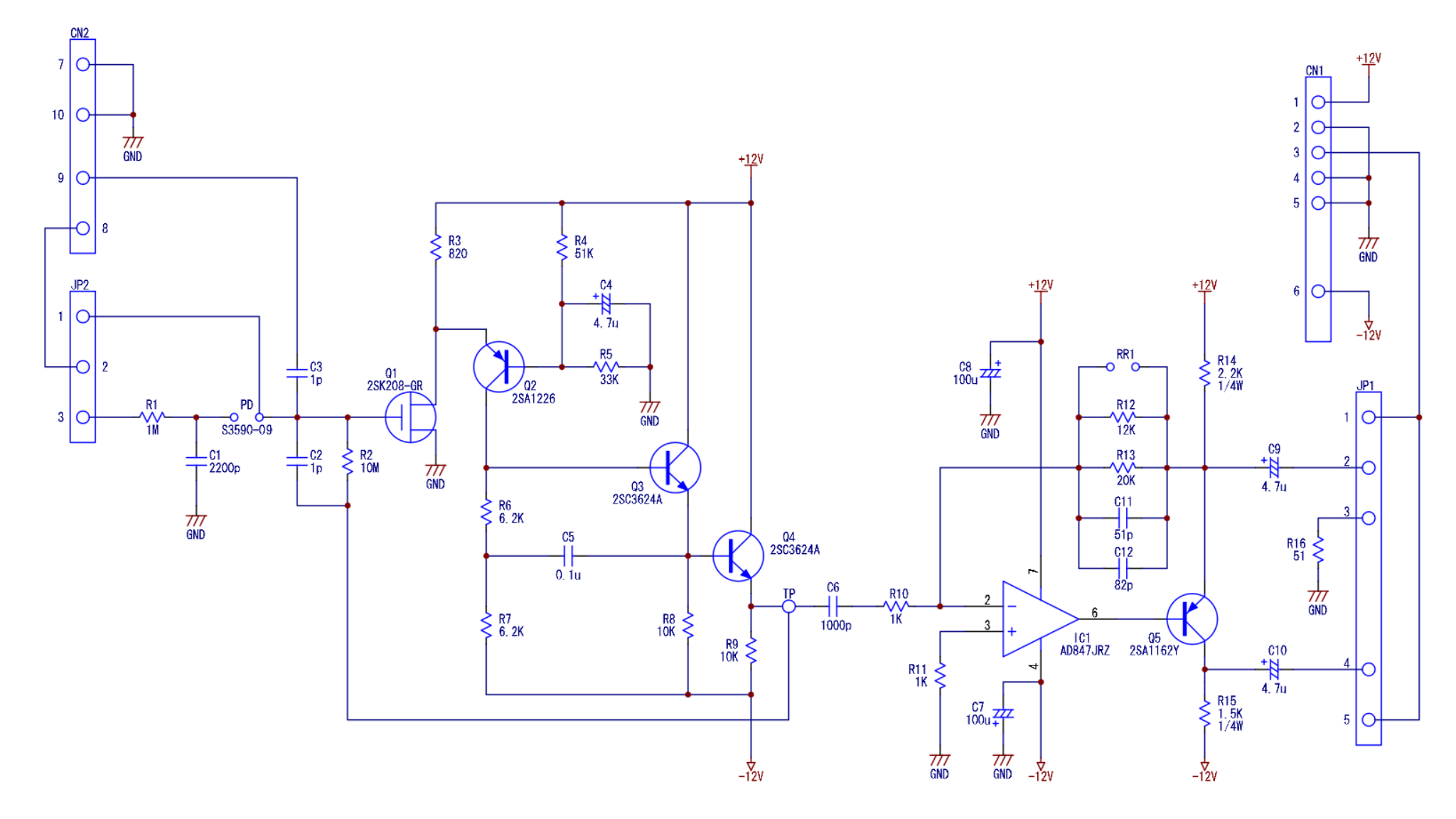}

	\caption{Circuit scheme of the HV supplier and amplifier circuit. The combination of resistors R12 and R13 leads to a $7.5$ gain. The $\pm12$V 
	input signal is connected to CN1 (1,5,6); the positive output signal is connected to CN1 (2,3); the HV is connected to CN2 (7,8).
	Connectors are set between JP1 (2,3), between JP1 (4,5), and between JP2 (2,3).
	\label{fig:PreAmp}}
\end{sidewaysfigure}

\begin{figure}
	\centering

	\includegraphics[width=.4\paperwidth]{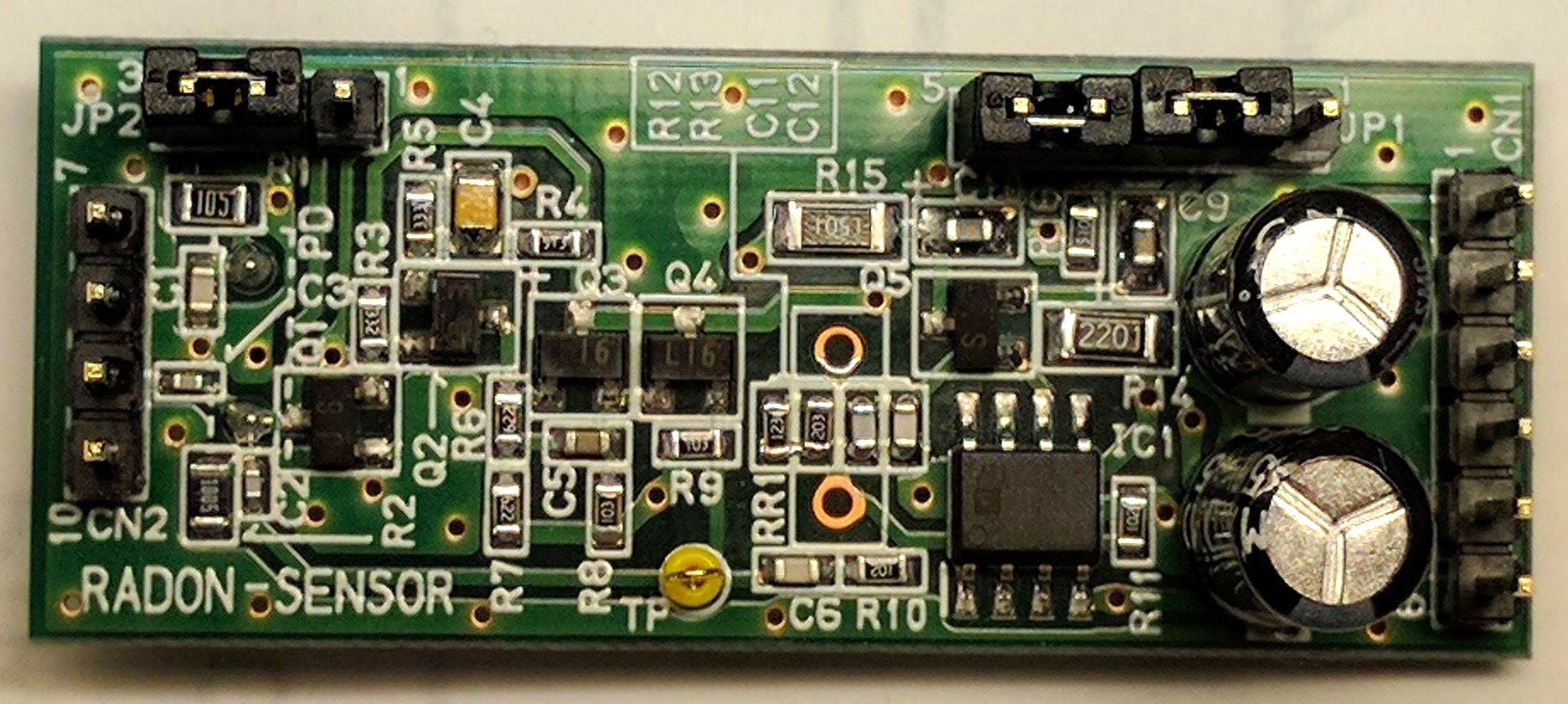}
	\caption{Picture of the amplifier circuit. View from above. Dimensions: 2 $\times$ 5 cm.
	\label{fig:PreAmpPic}}
\end{figure}

The HV supplier and the amplifier circuit are described in Fig. \ref{fig:PreAmp}, and a picture is shown in Fig. \ref{fig:PreAmpPic}. A HV of -120V is supplied to the whole detector. This
 bias voltage allows positive ions from the radon decays to be collected on the PIN photo-diode. The value of the HV was determined as the voltage for which the collection of $^{214}$Po and 
 $^{218}$Po positive ions is at its maximum.
 
The charge from the PIN photo-diode is converted to voltage pulses by the preamplifier circuit. The pulse is then shaped using AD847JRZ, inverted, and amplified. 
 The amplification gain of the amplifier circuit is $7.5$; this can be
reduced by adding a resistor in RR1 or increased by removing either R12 or R13. The signal is then sent to an analog-to-digital-converter (ADC) board 
specially developed to be used with Raspberry Pi systems. 

In order to measure the equivalent-noise-charge (ENC) from the preamplifier, the PIN photo-diode has been irradiated with $\gamma$-rays from an $^{241}$Am source. The 
59.4 keV photoelectric peak was measured, and allowed to measure the value of the ENC as 530 electrons (FWHM). 

\begin{sidewaysfigure}
	\centering
	\includegraphics[width=1.\paperwidth]{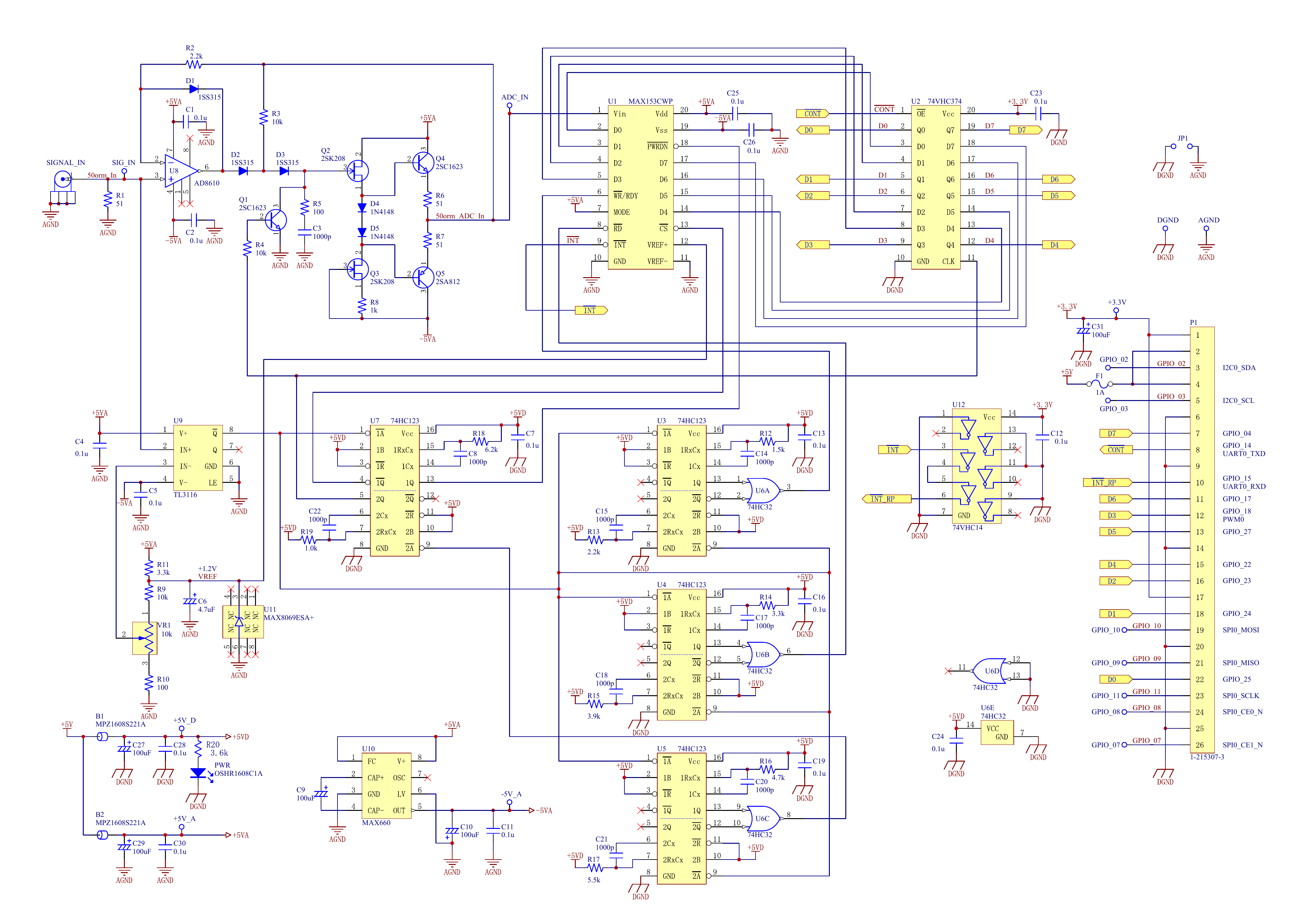}
	\caption{Circuit scheme of the ADC board used for the Raspberry Pi B, B+, 2B, and 3B systems.
	For Raspberry Pi B+, 2B, and 3B, the circuit also contains the circuit shown in Fig. \ref{fig:ENVCircuit}.
	\label{fig:ADCboard}}
\end{sidewaysfigure}

The ADC board circuit is described in Fig. \ref{fig:ADCboard}. It consists of a comparator circuit, a peak-holder circuit, an 8-bit ADC (MAX153CWP) \cite{Maxim}, 
a pulse timing circuit, and a latch circuit. It accepts an input signal via a 50$\Omega$-impedance LEMO connector with a maximal amplitude of 1000 mV.
This amplitude is converted into a channel number by the peak-holder circuit and the 8-bit ADC, and then sent to the Raspberry Pi.

Raspberry Pi B, B+, 2B, and 3B \cite{Raspberry} are currently used in the Kamioka mine. Due to the architectural differences between the different types of Raspberry Pi systems, 
two different versions of the ADC board have been developed. The first version, called ADC-B, is shown in Fig. \ref{fig:ADCB}. It is used with Raspberry Pi B, to which it is connected via
a 26-GPIO-pins connector. The second version, called ADC-B+, is shown in Fig. \ref{fig:ADCBplus}. It is used with Raspberry Pi B+, 2B, and 3B and is connected to the Raspberry 
Pi via a 40-GPIO-pins connector. The circuit of the ADC-B+ is similar to that of the ADC-B; however, because of the additional pins, 4 additional analog inputs are added to the 
circuit. These inputs accept signals with a maximal amplitude of 5000 mV and process them by a 4-channel 10-bit ADC (ADC104S021CIMM) \cite{Texas}. These additional inputs
are used to record environmental data (temperature, humidity, flow rate, etc.).

\begin{sidewaysfigure}
	\centering
	\includegraphics[width=.8\paperwidth]{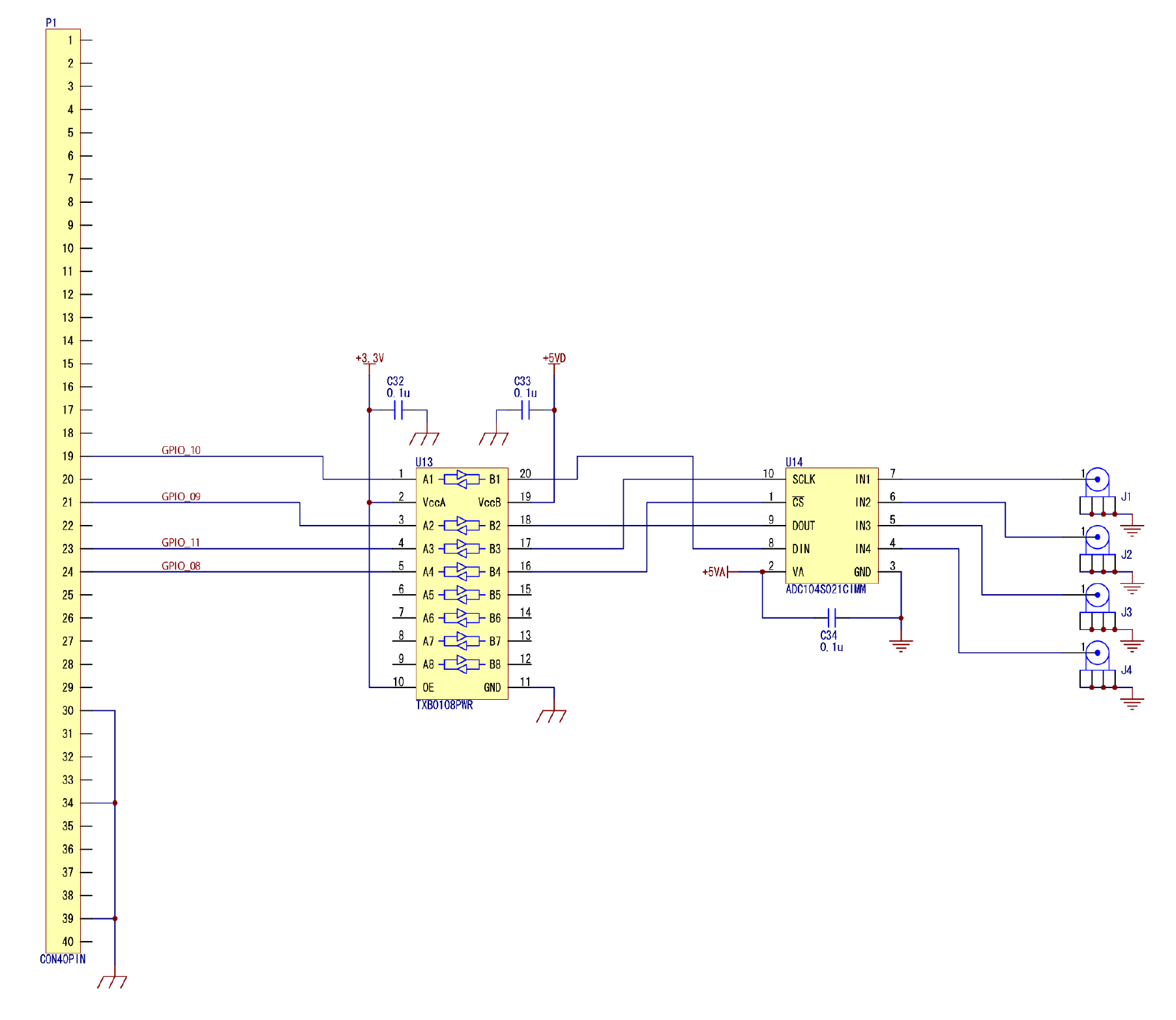}
	
	\caption{Additional part of the circuit scheme of the ADC board used for the Raspberry Pi B+, 2B, and 3B systems. \label{fig:ENVCircuit}}
\end{sidewaysfigure}

\begin{figure}
	\centering
	\includegraphics[width=.5\paperwidth]{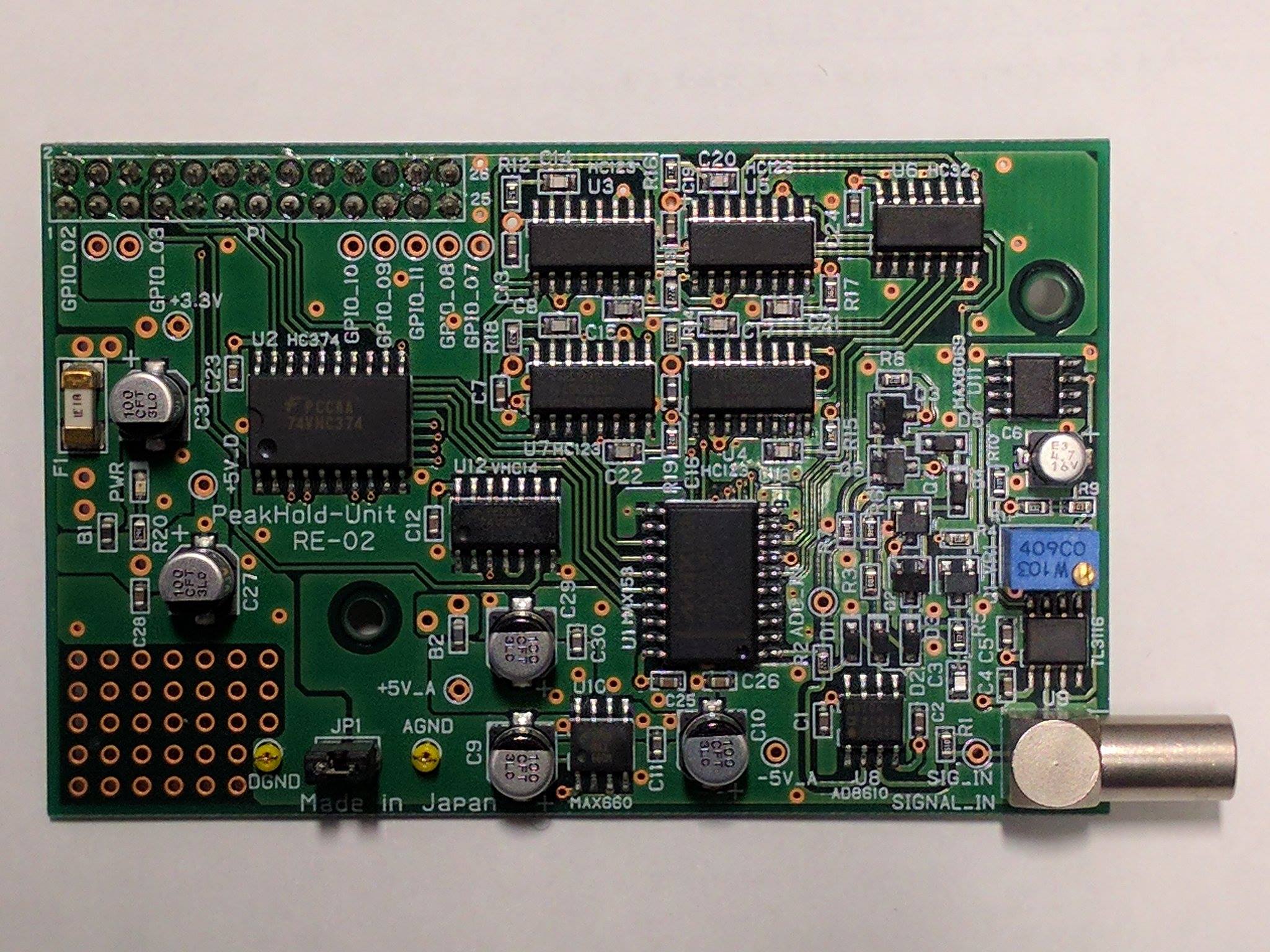}
	\caption{Picture of the ADC board version B, used for Raspberry Pi B systems. View from above. Dimensions: 5.5 $\times$ 8.5 cm.
	\label{fig:ADCB}}
\end{figure}

\begin{figure}
	\centering
	\includegraphics[width=.5\paperwidth]{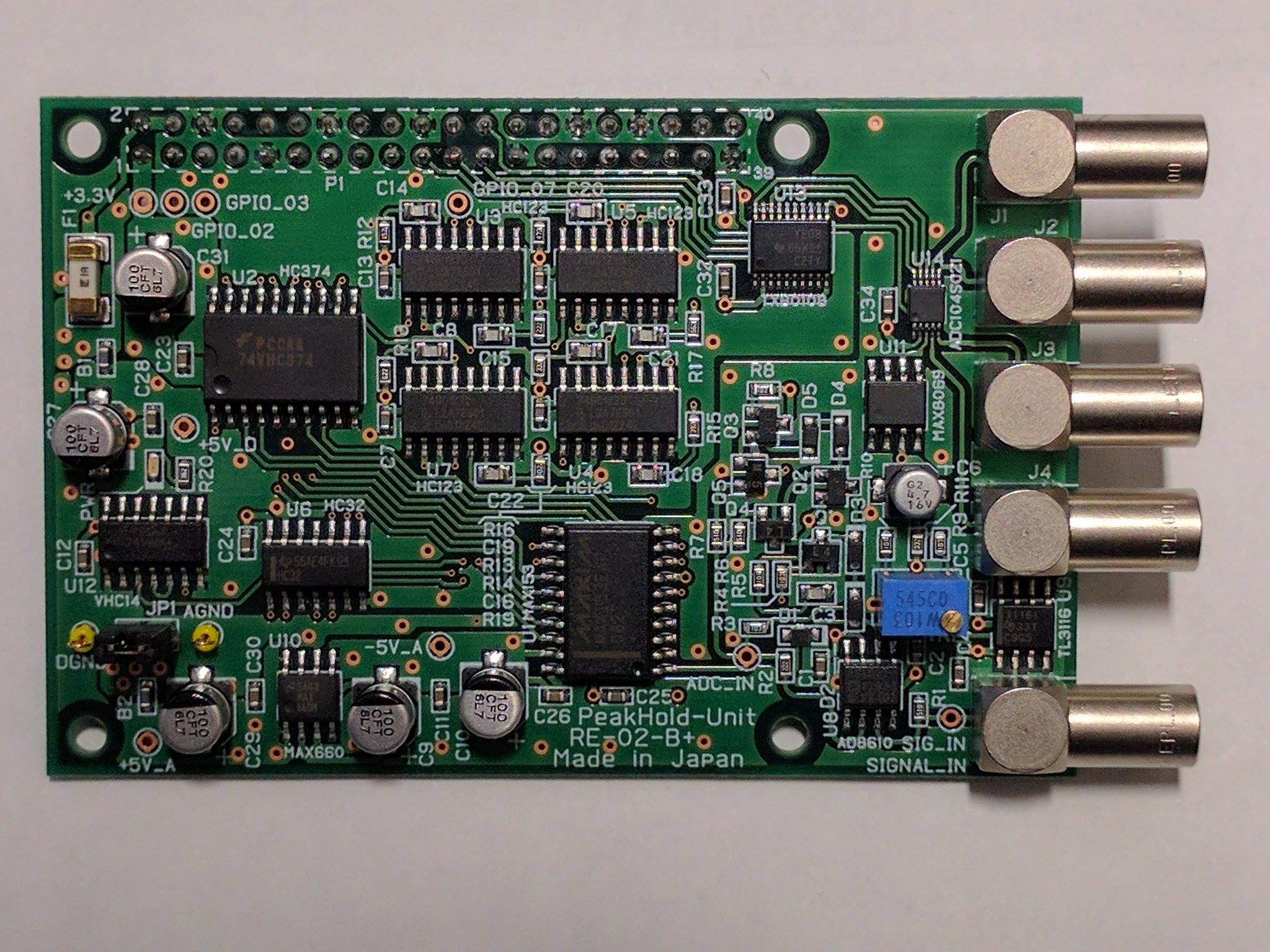}
	\caption{Picture of the ADC board version B+, used for Raspberry Pi B+, 2B, and 3B systems. View from above. Dimensions: 5.5 $\times$ 8.5 cm.
	\label{fig:ADCBplus}}
\end{figure}

The main part of the circuit is common to both versions and is described in Fig. \ref{fig:ADCboard}; 
the additional circuit for the 4 additional inputs is described in Fig. \ref{fig:ENVCircuit}. These boards are manufactured by WIN electronics Co., Ltd. \footnote{http://www.win-ei.com}.

\subsection{Data acquisition}

The ADC board is monitored by software to perform data acquisition. This software was
developed using C++ and uses the wiringPi framework \cite{wiringPi} to communicate with the ADC board through the GPIO-pins connector.

This software was developed to record data with fixed time interval measurements. However, the CPU speed on the Raspberry Pi systems limits the minimal interval which can be used. 
To estimate this minimal time interval, hardware-limitation tests have been performed using a Raspberry Pi 3B equipped with a 48-MB/s TOSHIBA 32-GB micro-SD card. 
Tab. \ref{tab:MinimalTime} summarizes the number of times the actual interval was different than the expected interval.

\begin{table}[hbt!]
	\small
	\centering
	\begin{tabular}{| c | c | c | c  | }
		\hline
		Expected	& Actual mean		& \multicolumn{2}{|c|}{Different intervals} 		\\
		 interval	&  interval 		& $\%$ 				& $\#$ per day		\\
		\hline
		$1$ sec 	& $1.59 \pm 0.50$ sec 	& $59.6201 \pm 1.1182 \%$	& $51806 \pm 967$	\\
		$2$ sec 	& $2.01 \pm 0.21$ sec 	&  $0.2717 \pm 0.0008 \%$	&   $117 \pm 1$		\\ % 3 days
		$3$ sec 	& $3.01 \pm 0.15$ sec 	&  $0.1292 \pm 0.0005 \%$	&    $37 \pm 1$		\\ 
		$4$ sec 	& $4.00 \pm 0.13$ sec 	&  $0.0356 \pm 0.0002 \%$	&  $8 \pm 0$ ($0.05$)	\\ % 0.0355594 percent 0.000131506
		$5$ sec		& $5.00 \pm 0.10$ sec 	&  $0.0296 \pm 0.0002 \%$	&  $5 \pm 0$ ($0.02$)	\\ % 0.0295743 percent 0.000127149
		\hline
	\end{tabular}

	\caption{Summary of the study of the minimal time interval measurement. The ``Different intervals'' columns show the ratio (third column) and number of times per day (fourth column) 
	that the actual interval was not the same than the expected interval. Only statistical uncertainties are shown. 
	\label{tab:MinimalTime}}
\end{table}

From these results, we can see that the CPU speed of the Raspberry Pi cannot perform 1 measurement per second as this configuration leads to obtaining less data than when asking
for 1 measurement every 2 seconds. Hence, we determined that the minimal interval that can be used without critical data loss is \textit{2} seconds. 

Additional tests showed that electronics are able to support without lost an input of $\sim23$kHz in case of Raspberry Pi B and B+, and of $\sim28$kHz in case Raspberry Pi 2B and 3B. 
At this rate, the current version of the software requires to be run with a 4 sec (or less) interval in order to avoid data loss. 
Taking into account the calibration factor showed later, and assuming a humidity of $10$ $\mathrm{g/m}^{3}$, this rate theoretically corresponds to a maximal radon concentration of $\sim260$ $\mathrm{MBq/m}^{3}$ in case of 
Raspberry Pi B and B+, and of $\sim315$ $\mathrm{MBq/m}^{3}$ in case of Raspberry Pi 2B and 3B. 

%The ADC board dependency of the measurement was checked by deploying one detector connected with two different RaspberryPis via a LEMO-T connector. 

\section{Calibration of the radon detector}
\label{sec:Calibration}

\begin{figure}

	\centering
	\includegraphics[width=.7\linewidth]{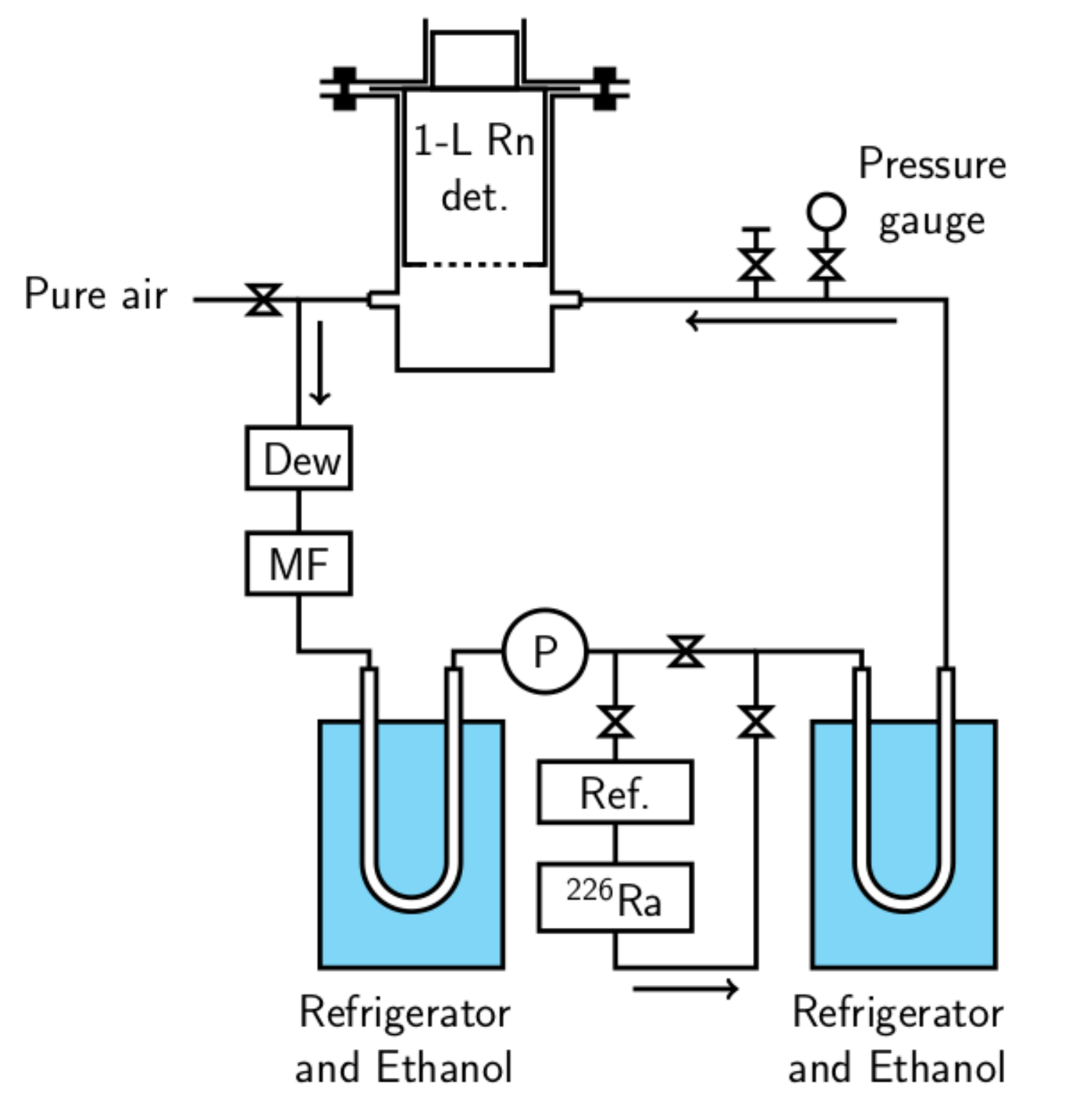}
	
	\caption{Schematic of the 1-L Rn detector calibration system. ``Dew'' indicates the position of the humidity sensor, ``MF'' indicates the position of the Mass Flow controller used to 
	monitor the air flow in the loop, ``P'' indicates the position of the air pump, ``Ref'' indicates the position of the sampling point used to measure the actual radon concentration, 
	and ``$^{226}$Ra'' indicates the position of the $^{226}$Ra source. The arrows show the direction of the air flow.
	\label{fig:CalibSys}}
\end{figure}

Previous studies showed that the $^{222}$Rn daughter nuclei can be collected and neutralized by water molecules in the air \cite{ChuNeutralisation}. Therefore,
the detection efficiency depends on the absolute humidity in the environment. A study of the humidity dependence of the calibration factor $C_{F}$ (in $\mathrm{(counts/day)/(Bq/m^3)}$)
was thus performed. The calibration system was a closed loop including a 1-L detector, a humidity meter, a radon source, and an air pump. A refrigerator was used
to control the dew-point temperature of the system. A schematic of the calibration system is shown in Fig. \ref{fig:CalibSys}.
The calibration was performed in two different periods. At first, we measured the high humidity dependence ($>6$ $\mathrm{g/m^3}$); 
then, we measured the low-humidity dependence ($<1$ $\mathrm{g/m^3}$). 
The humidity was measured with a MICHELL sensor (PCMini52) during the first campaign and with a VAISALA Dew Point meter (DMT340) during the second. The MICHELL sensor
measured the relative humidity and air temperature, whereas the VAISALA sensor measured the dew point and the air temperature. The conversion to absolute
humidity was performed using the VAISALA-humidity formula in the $[-20,50]$\,\textcelsius\,approximation \cite{VAISALAdoc}. The actual radon concentration was measured 
with a Pylon scintillator counter (Lucas cell 300A) \cite{PYLON}. The accuracy of the Pylon scintillator counter has been estimated to be $\pm4\%$,
this value is used as a systematic error in the calibration.

The peak from $^{218}$Po $\alpha$ decay can be overlapped with the peak from $^{216}$Po $\alpha$ decay in the Th-series-decay chains. Hence, the number of counts used 
to compute the radon concentration is the integral of the peak from the $^{214}$Po $\alpha$ decays. This approach was also used in \cite{10.1093/ptep/ptv018}.

\begin{figure}
	\centering
	\includegraphics[width=.5\paperwidth]{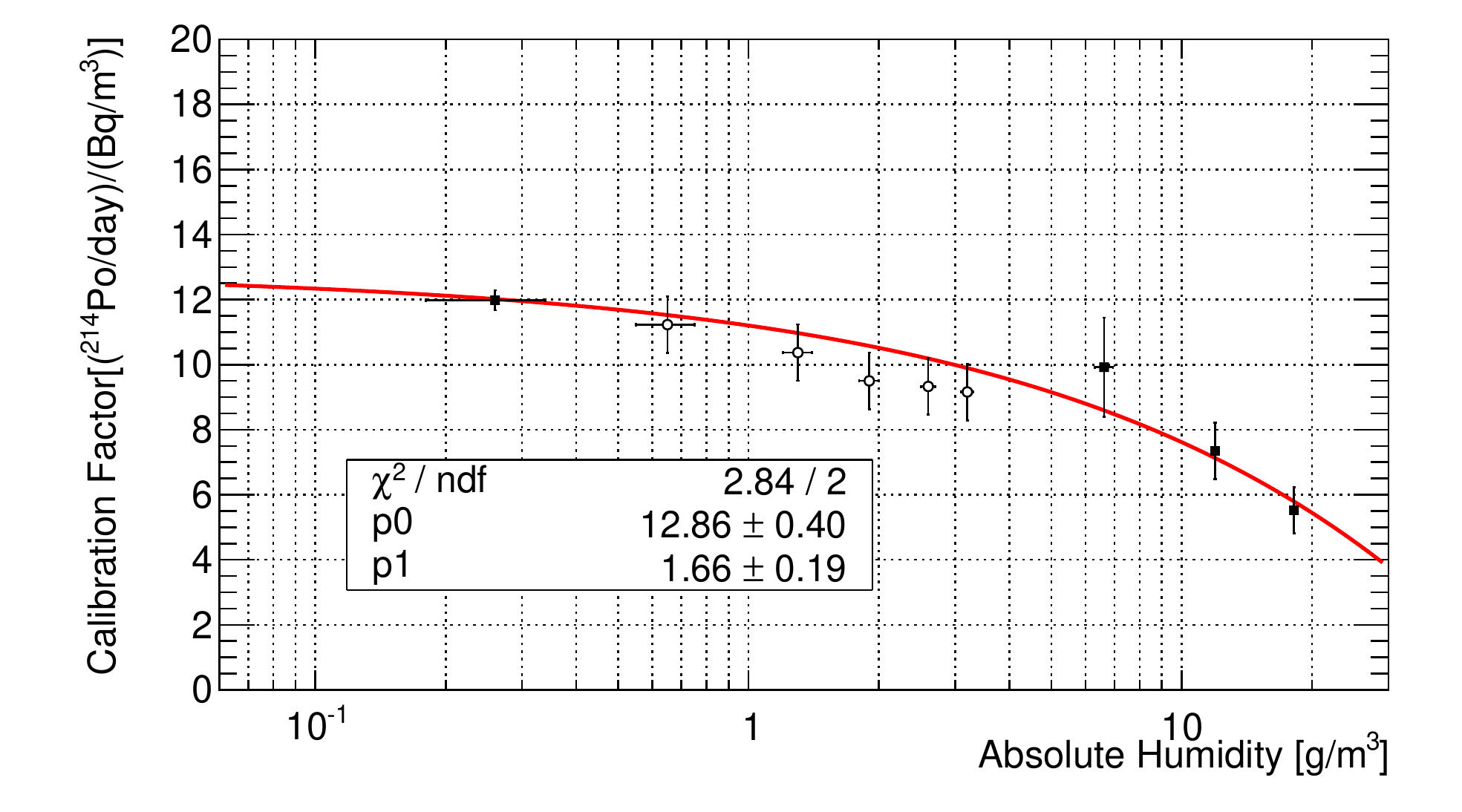}
	\caption{1-L detector-calibration factor in $\mathrm{(counts/day)/(Bq/m^3)}$ as a function of the absolute humidity. The uncertainties 
	of the absolute humidity for the two last points are $0.06$ $\mathrm{g/m^3}$ and $0.12$ $\mathrm{g/m^3}$ (from left to right) and are therefore not visible here.
	The red solid line shows the fit with a function $f(A_{H}) = p0 - p1 \times \sqrt{A_{H}}$, with $A_{H}$ being the absolute humidity and $p0$ and $p1$ being the function's parameters. 
	%The blue surface between $6$ $g/m^3$ and $11$ $g/m^3$ indicates the typical range of humidity in the Kamioka Mine.
	The white markers refer to the data points from \cite{DevSmall}. These data points were not used to determine the fit function.
	\label{fig:fitCalibration}}
\end{figure}

Fig. \ref{fig:fitCalibration} shows the measured calibration factors in $\mathrm{(counts/day)/(Bq/m^3)}$ as a function of the absolute humidity. The data were found to be
best described by $f(x) = p0 - p1 \times \sqrt{x}$, as in \cite{10.1093/ptep/ptv018}. The fit result is shown as:

\begin{equation}
	C_{F} (A_{H}) = (12.86 \pm 0.40) - (1.66 \pm 0.19) \sqrt{A_{H}},
\end{equation}

with $C_{F}$ being the calibration factor in $\mathrm{(counts/day)/(Bq/m^3)}$ and $A_{H}$ being the absolute humidity in $\mathrm{g/m^3}$. 
The uncertainties include both statistical and systematics errors. This fit result is consistent with the previous study 
performed in \cite{DevSmall}, as is shown in Fig. \ref{fig:fitCalibration} with the data points from Fig. 6 in \cite{DevSmall}.

In order to measure the background level of 1-L detector, we sealed a 1-L detector in a stainless steel container purged with pure air.
The background level was measured during one month to be $0.65\pm0.15$ $\alpha_{3}$ count/day. This allowed to determine the detection lower limit 
with the Currie's method \cite{CurrieMethod} as $\sim0.4$ $\mathrm{Bq/m^{3}}$ for a $10$ $\mathrm{g/m^{3}}$ humidity, for one day of measurement. 

\begin{figure}

	\centering
	\includegraphics[width=.7\linewidth]{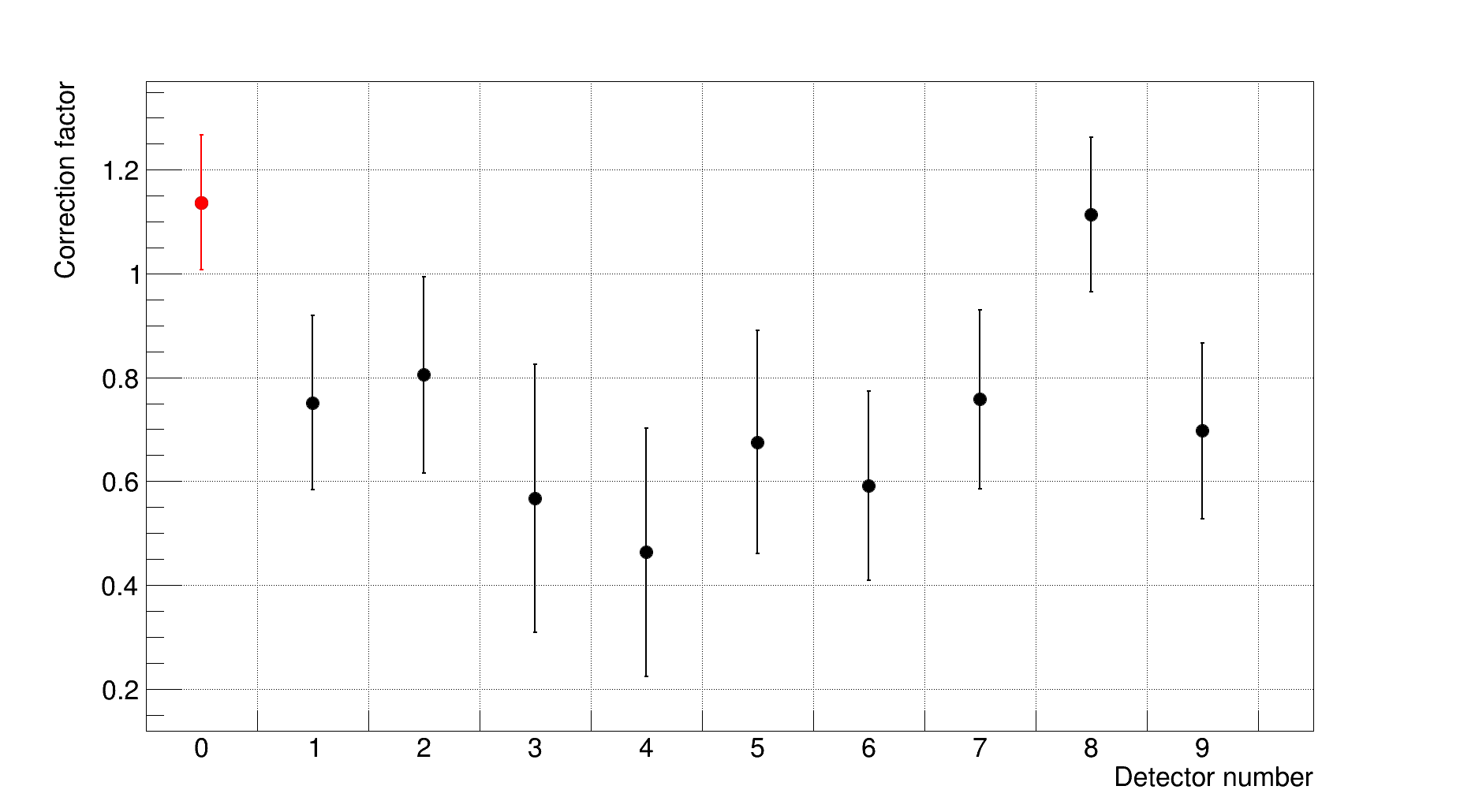}
	
	\caption{Correction factor for the detectors used in the Results part of this paper. The x-axis show the detector number. The correction factor for the reference
	detector is shown in red (detector \#0).  
	\label{fig:CorrectionFactor}}
\end{figure}

Due to the differences in the HV feed-through manual manufacturings and in the PIN photodiode settings, the detector response is slightly different for each detectors. 
In order to take this into account, each detector was individually calibrated. This calibration was done in a high radon concentration environment ($>500$ $\mathrm{Bq/m^{3}}$ in the mine) 
where each detector was deployed together with a reference detector (ionization chamber detector, SAPHYMO Alpha
GUARD PQ2000) for several days. The stability of this calibration was checked with the deployment of a ``reference''
detector over the full calibration period. The relative deviation between the Alpha-Guard detector and a 1-L detector is used to determine a correction factor, 
which is applied on the detector result. The correction factors used for the detectors presented in this paper are shown in Fig. \ref{fig:CorrectionFactor}. 
The error of the measured correction factor is dominated by the statistics due to the relatively short measurement period for quick deployment of $\sim10$ detectors.

\section{Data analysis}

\begin{figure}
	\centering
	\includegraphics[width=.5\paperwidth]{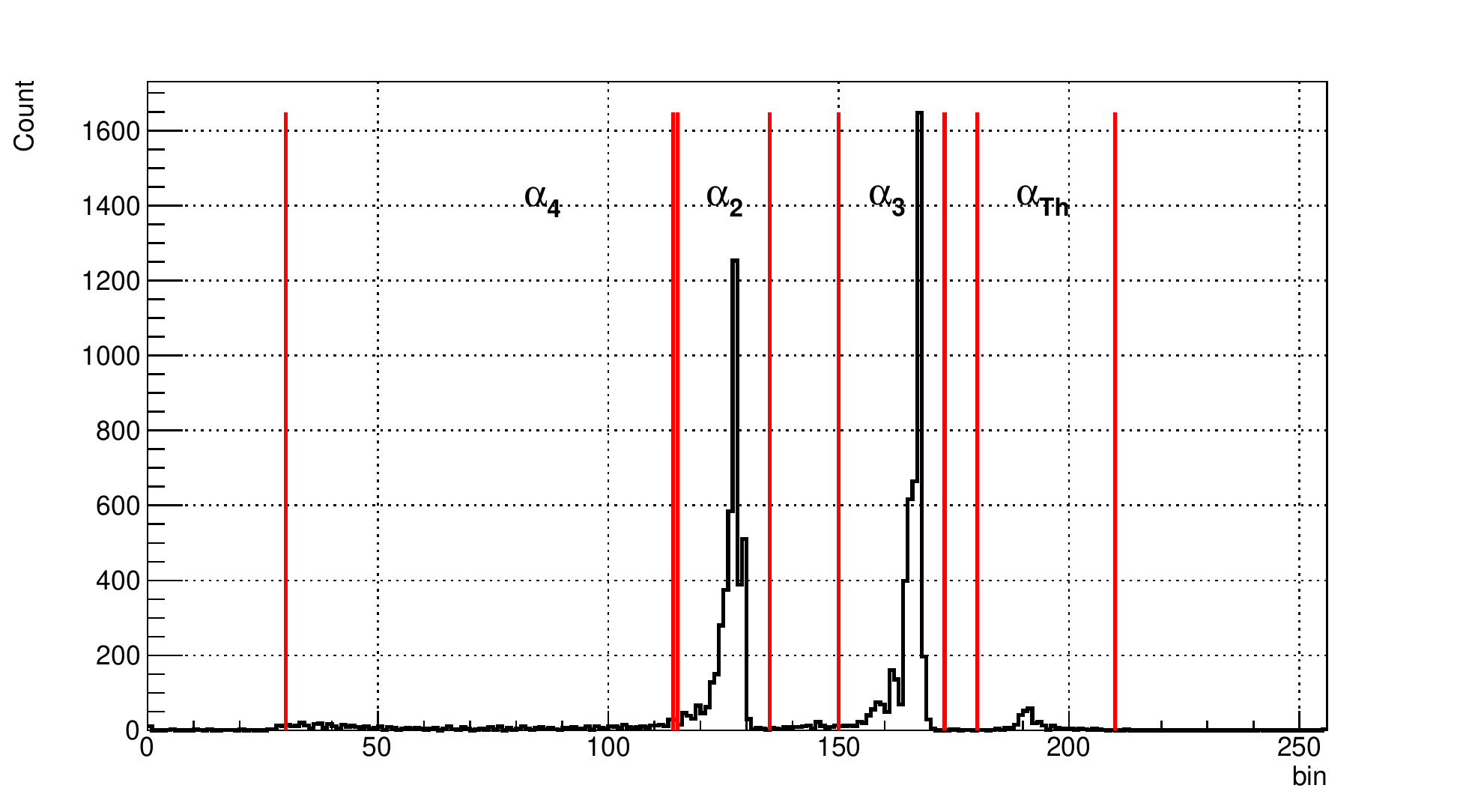}
	\caption{Spectrum from one 1-L radon detector. The red lines show the cuts used to determine the peak positions.
	This detector is using a brand-new PIN photo-diode (less than one month exposure), hence the absence of $^{210}$Pb peak ($\alpha_{4}$).
	\label{fig:RnSpectrumCut}}
\end{figure}

A framework, based on C++ and on ROOT/CERN \cite{ROOTCERN}, was developed to automatically analyze the data from many different radon detectors in the Kamioka mine. 
Using a standard approach from \cite{KamiokaRadonPaper}, 
the spectrum obtained from the ADC board is analyzed. Cuts are defined to integrate the spectrum between 4 intervals,  of which 3 correspond to the main $\alpha$
decays from the $^{222}$Rn-decay chain ($\alpha_2$, $\alpha_3$, and $\alpha_4$) and the last to the  $\alpha$ 
decay of $^{212}$Po ($E(\alpha_{Th}) =8.784$ MeV) in the $^{232}$Th-series-decay chains. The measurement of the number of counts from $^{212}$Po $\alpha$ decays allows estimation of 
 the contamination from $^{216}$Po $\alpha$ decays (E = 6.788 MeV) in the $^{218}$Po $\alpha$-decay peak, as explained in Section \ref{sec:Calibration}.  $\alpha$ 
particles from $^{212}$Po are denoted by $\alpha_{Th}$ in the following. The cuts are defined separately 
 for each detector because small variations of the gain between each amplifier can lead to variations in the peak positions. Fig. \ref{fig:RnSpectrumCut} shows an example
 of the cuts performed for one detector.
 
The main difficulties with the automatic analysis of dozens of detectors is the possible fluctuation of the peak positions. Each time a detector HV is restarted or the PIN photo-diode
or amplifier is replaced, the gain can vary, leading to a change of the peak position. 
The software is thus able to determine whether the peak positions change and raises an error if this happens. Human action is then needed to determine the new peak positions. 
In normal operation, this maintenance is not needed for more than 2 years. This is mainly needed after power failure in the mine, and if the PIN photo-diode or the amplifier is replaced.

Data analyses are performed on a server in the Kamioka network. However, thanks to their higher performances, data analyses can also be performed directly on the Raspberry Pi 2B 
and 3B systems. This allows the use of these electronics as standalone units outside of the Kamioka network.

\section{Results}

\begin{figure}
	\centering
	\includegraphics[width=.55\paperwidth]{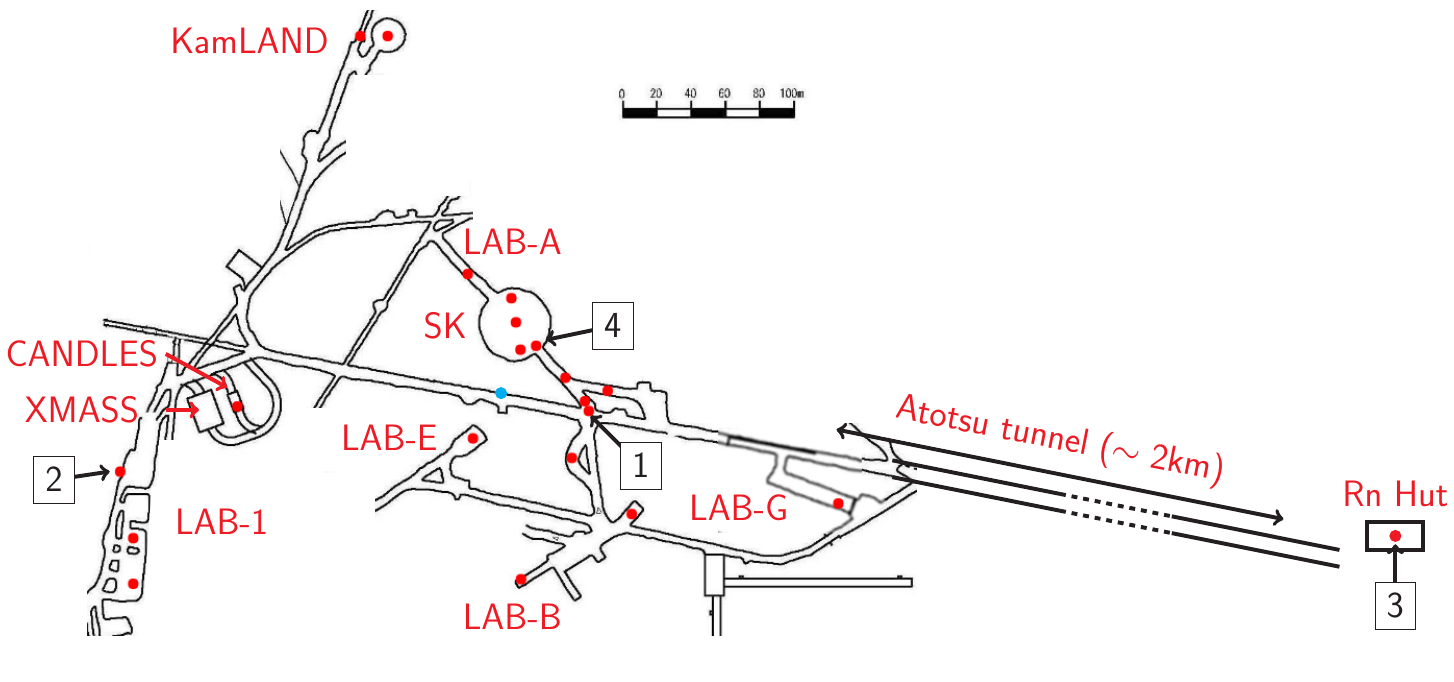}
	\caption{Map of the experimental area of the Kamioka mine. The red points refer to the detectors' locations. The numbers indicate the devices used in this article. The blue point refers to the anemometer's position.
	\label{fig:KamiokaMap}}
\end{figure}

\begin{figure}
	\centering
	\includegraphics[width=.25\paperwidth]{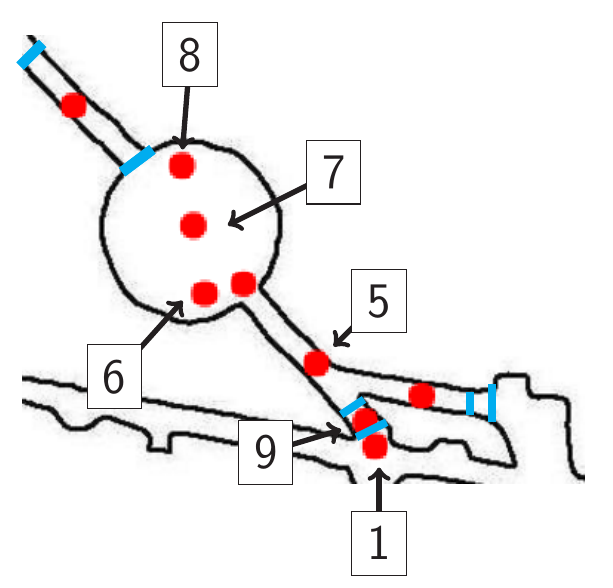}
	\caption{Zoom in on the Super-Kamiokande experimental area. The red points indicate the detectors' locations. The numbers are an indication for the detectors used in this article, with number 1 showing
	a reference from Fig. \ref{fig:KamiokaMap}. The blue lines indicate the airtight doors' positions.
	\label{fig:KamiokaZoom}}
\end{figure}

Twenty one 1-L detectors using Raspberry Pi electronics are monitoring the radon concentration in the air at various positions in the Kamioka mine. In this section, the measurements from a selection of these
detectors are shown and discussed. Fig. \ref{fig:KamiokaMap} shows the positions of these twenty one detectors in the Kamioka mine, and Fig. \ref{fig:KamiokaZoom} zooms in on the Super-Kamiokande
area.

\subsection{Radon concentration in the mine}

\begin{figure}
	\centering
	%\begin{subfigure}{\linewidth}
	%	\centering
	%	\includegraphics[width=\linewidth]{Detect25}
	%	\caption{Atotsu tunnel, front of the SK area (detector \#1)} \label{fig:sfig25}
	%\end{subfigure}%
	
	%\begin{subfigure}{\linewidth}
	%	\centering
	%	\includegraphics[width=\linewidth]{Detect14}
	%	\caption{XMASS 2nd pure-water system room (detector \#2)} \label{fig:sfig14}
	%\end{subfigure}
	\includegraphics[width=.9\linewidth]{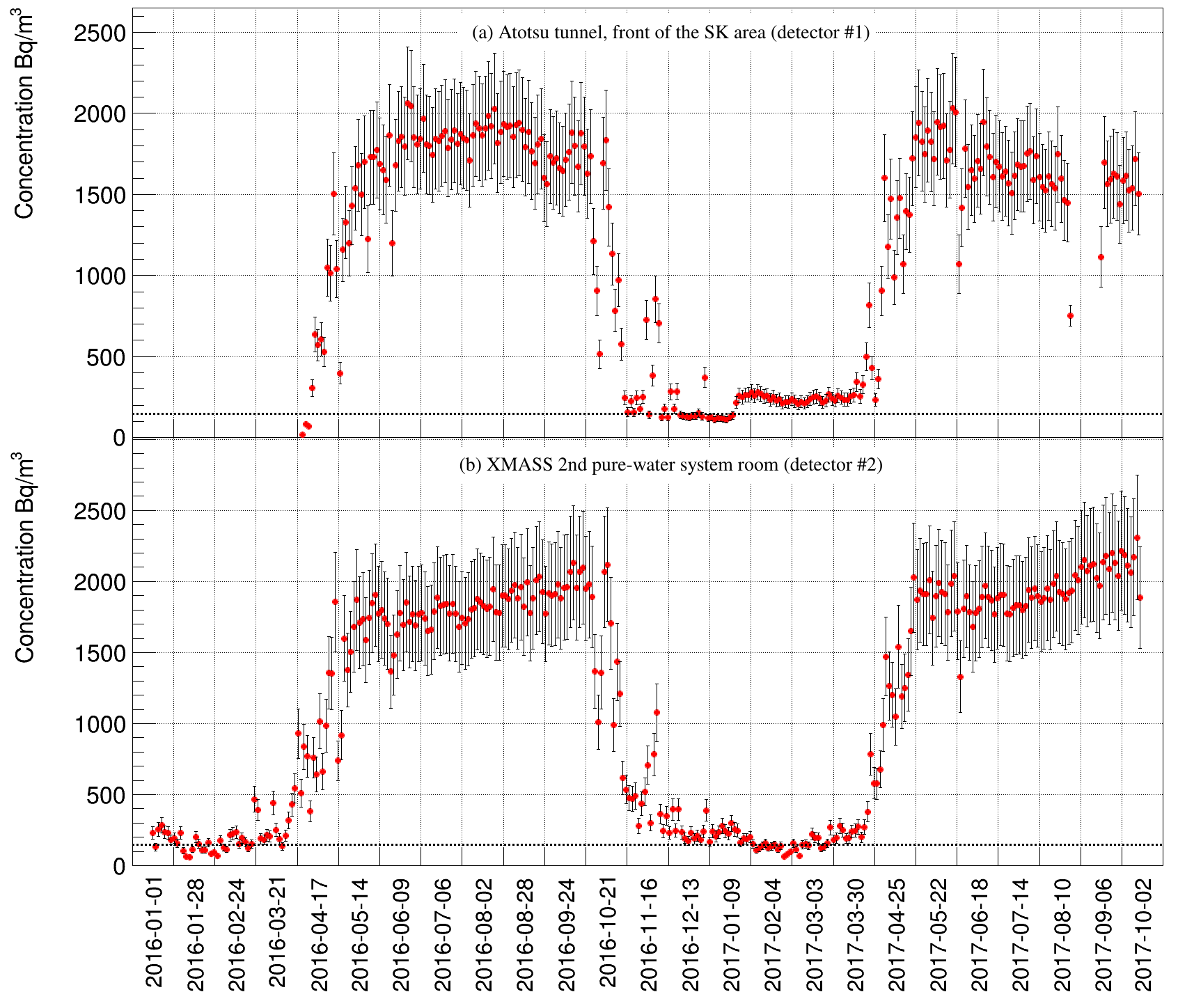}
	
	\caption{Radon concentration (2-days-averaged) measured in Kamioka's mine areas without fresh-air input.
	The dotted lines indicate the 148-Bq/m$^3$ U.S. limit for indoor radon concentration \cite{RadonRisk}. The top figure (Atotsu tunnel) starts
	around April 7th 2016 as the detector was not deployed in the mine beforehand. The error bars include statistical uncertainties, as well as 
	the uncertainty on the detector's correction factor.
	\label{fig:TunnelConcentrations}}
\end{figure}

\begin{figure}

	\centering
	\includegraphics[width=.8\linewidth]{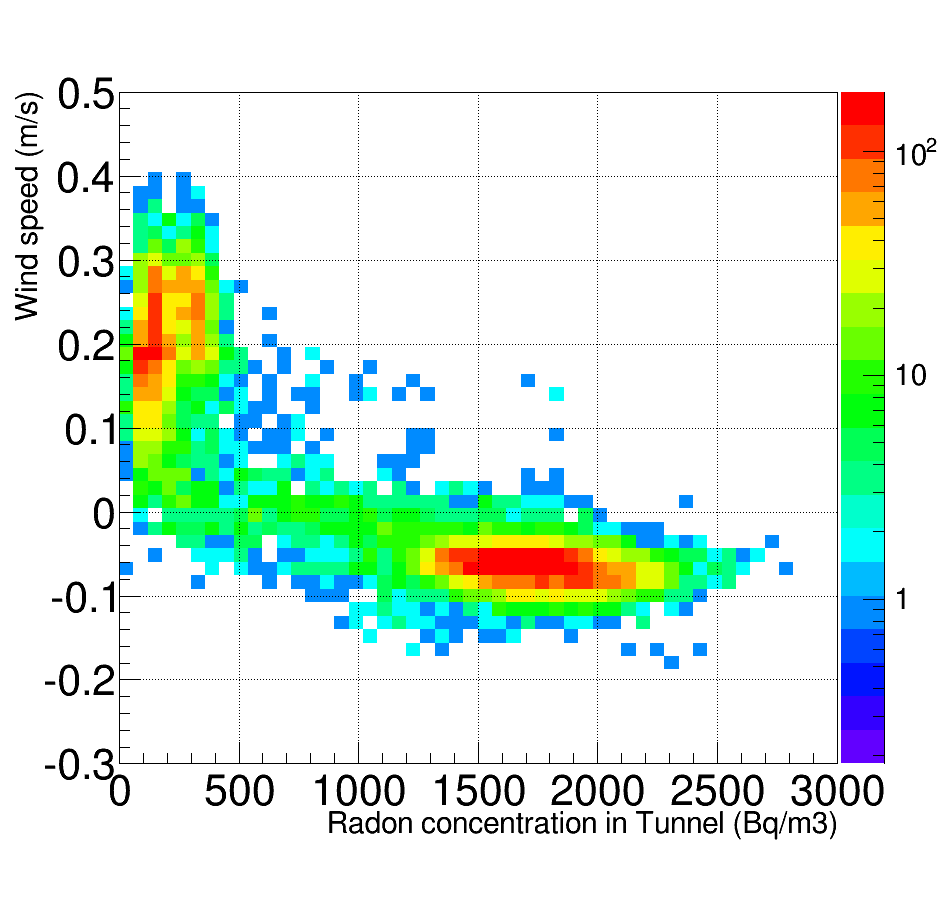}
	
	\caption{Correlation between the hourly-averaged radon concentration measured in the Atotsu tunnel and the speed of the wind in this tunnel. Positive values indicate 
	an air flow coming from outside of the mine and negative values indicate the opposite. The color scale shows the number of measurements in each bin.
	\label{fig:TunnelCorr}}
\end{figure}

Fig. \ref{fig:TunnelConcentrations} shows the radon concentrations from January 2016 to October 2017 in the tunnels of the mine 
from the detectors \#1 and \#2 in Fig. \ref{fig:KamiokaMap} (directly exposed to the air of the mine).
The humidity in these areas ranges from 8 $g/m^{3}$ to 12 $g/m^{3}$. A clear seasonal variation of the radon concentration is 
observed. In the approximate period between April to November, the radon concentration is high ($>1000$ $Bq/m^3$), whereas it is low ($<500$ $Bq/m^3$) between November and April.
In the Atotsu tunnel, the main access tunnel to the Super-Kamiokande area, we measure the speed and direction of the wind with an anemometer, whose position is indicated in Fig. \ref{fig:KamiokaMap}.
A correlation is observed between the radon concentration in the Atotsu tunnel and the direction of the wind, as shown in Fig. \ref{fig:TunnelCorr}. The radon
concentration in the Atotsu tunnel is high when the wind comes from inside the mine and low when it comes from outside. 
Our interpretation is that fresh air with low radon concentration is brought into the mine when wind is coming in from outside, whereas the high-radon-concentration air is pushed outside
the mine when the wind is blowing in the opposite direction. An increase of the radon concentration is observed from 2017, January 10th. This increases is due to a temporary modification 
of the water flow in the mine, leading radon-rich water near the detector. 

\begin{figure}
	\centering
	\includegraphics[width=.9\linewidth]{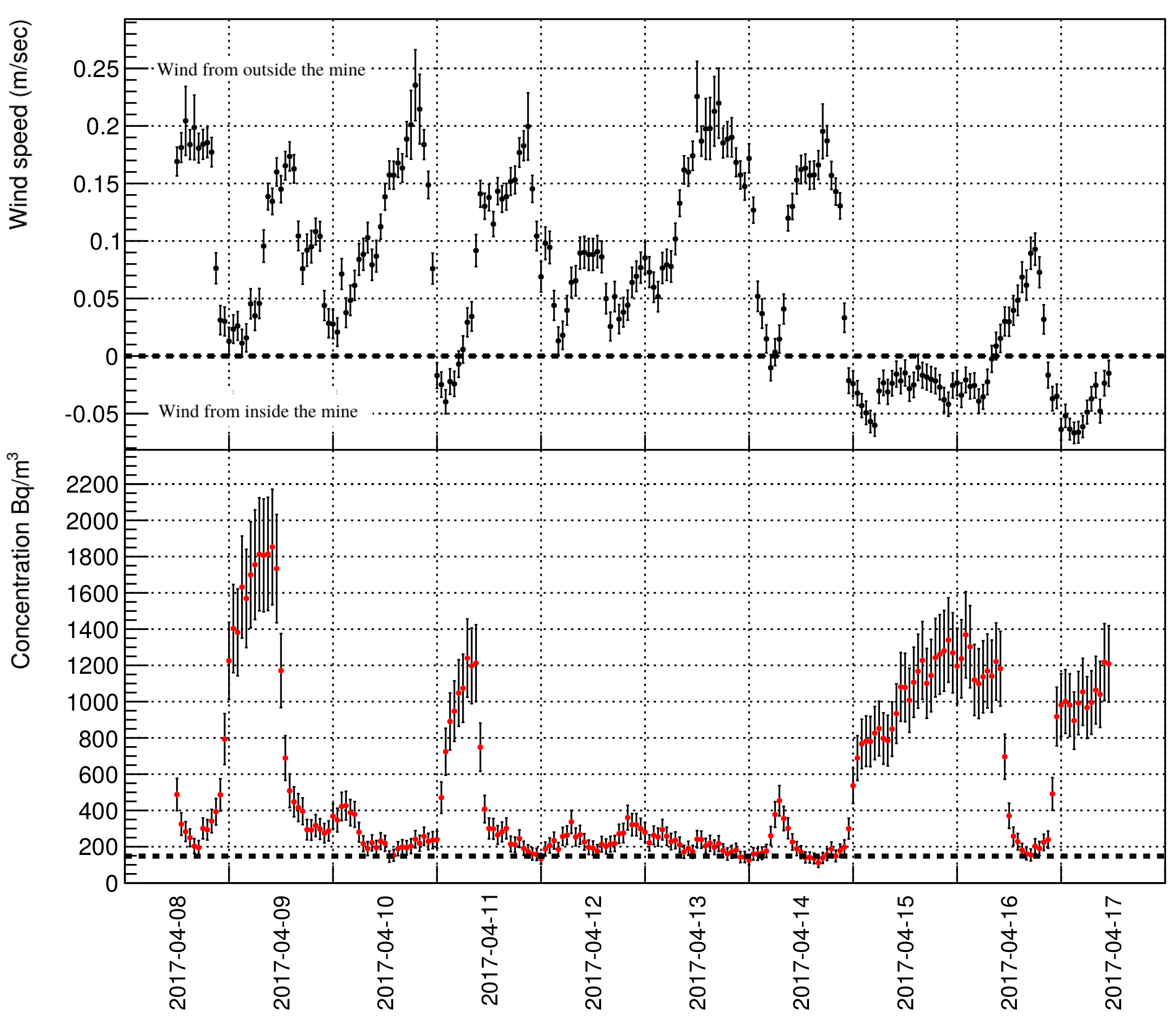}
	
	\caption{Hourly-averaged radon concentration (bottom) and wind velocity (top) measured in the Atotsu tunnel (detector \#1) from April 8th to April 16th.
	\label{fig:TunnelWind}}
\end{figure}

In Fall (October-November) and Spring (April), daily fluctuations of the radon concentration are observed. Fig. \ref{fig:TunnelWind} illustrates this phenomenon.
The radon concentration typically increases during daytime (usually from noon) and decreases during nighttime. Our interpretation of this phenomenon is that the temperature 
outside the mine fluctuates around the value leading to a change of the wind direction in the tunnel.

\subsection{Radon concentration in the fresh air}

As shown in the previous section, the radon concentration in the air in the mine is very high. Therefore, to reduce the exposure of workers and 
experiments, fresh air needs to be brought into the experimental areas. This fresh air is brought through a 1.8-km pipe \cite{FreshAirPaper} going through 
the Atotsu tunnel. This pipe brings $100$ m$^3$ of fresh air per minute into the mine. The humidity in the pipe is kept between $2.5$ and $7$ $g/m^{3}$ by a dryer.

\begin{figure}
	\centering

	%\begin{subfigure}{\linewidth}
	%	\centering

	%	\includegraphics[width=\linewidth]{Detect33}
	%	\caption{Fresh air (input) (detector \#3)} \label{fig:sfig33}
	%\end{subfigure}%
	
	%\begin{subfigure}{\linewidth}
	%	\centering
	%	\includegraphics[width=\linewidth]{Detect32}
	%	\caption{Fresh air (output in Super-K dome) (detector \#4)} \label{fig:sfig32}
	%\end{subfigure}%
	\includegraphics[width=.9\linewidth]{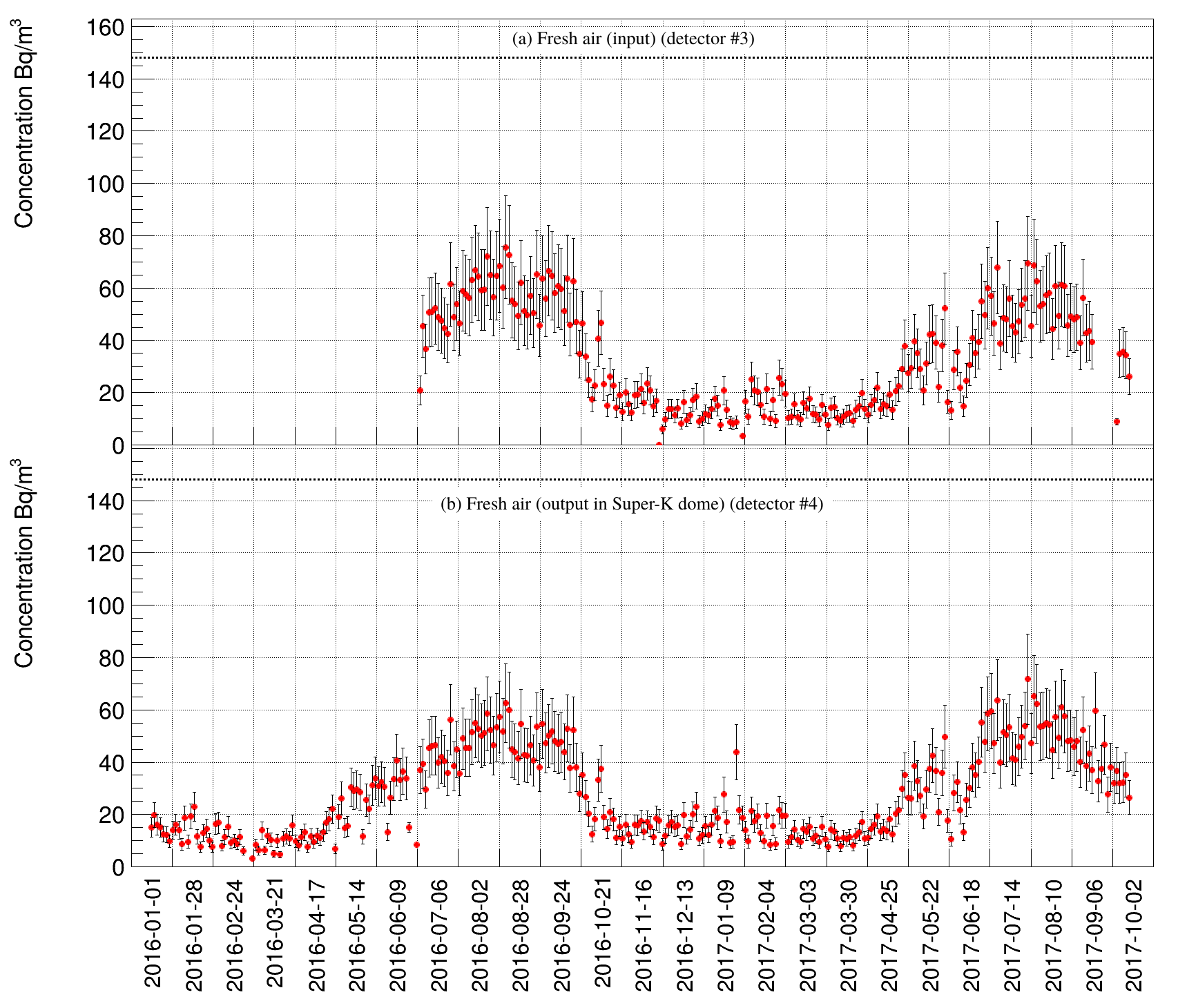}

	\caption{2-days-averaged radon concentration measured at the input (top) and output (bottom) of the fresh-air pipe. The detector at the input of 
	the fresh-air pipe was deployed at the end of June 2016, hence the absence of data beforehand. The error bars include statistical uncertainties, as well as 
	the uncertainty on the detector's correction factor.
	\label{fig:DuctConcentrations}}
\end{figure}

\begin{figure}

	\centering
	\includegraphics[width=\linewidth]{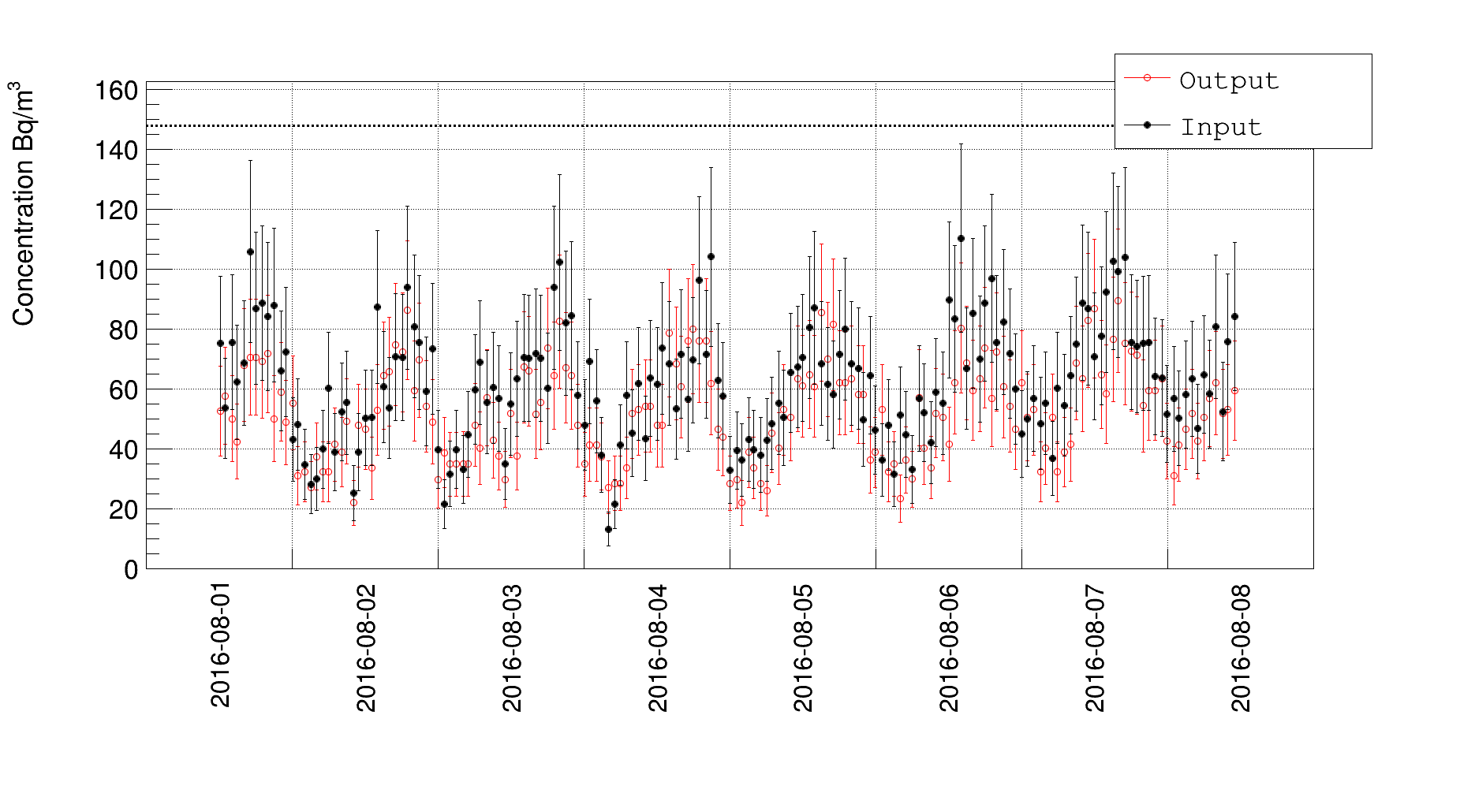}

	\caption{Hourly-averaged radon concentration measured at the input (plain circles, black) and output (red) of the fresh-air pipe.
	\label{fig:DuctHourly}}
\end{figure}

We positioned two detectors to monitor the radon concentration in this pipe. One was placed at the input outside of the mine, in the so-called ``Rn hut'' that houses 
a large compressor used to send the fresh air, and the other detector was placed at one of the outputs of the pipe in the Super-Kamiokande dome. 
Fig. \ref{fig:DuctConcentrations} shows the concentrations measured by these 
detectors from January 2016 to October 2017. A seasonal variation is observed on this figure, with an averaged radon concentration of $50.7 \pm 1.0$ $Bq/m^{3}$ during in summer, 
and $ 19.7 \pm 0.2$ $Bq/m^{3}$ during the rest of the year. This observation suggests a feed back of the high radon concentration air from Atotsu tunnel. 
Fig. \ref{fig:DuctHourly} shows the hourly-averaged radon concentration for both detectors over August 2016 and suggests that both detectors measured the same 
concentration and that there are no leaks in the pipe (i.e., no additional radon coming from the air in the mine).

Fig. \ref{fig:DuctHourly} also indicates variations between daytime and nighttime radon concentrations in the fresh air. Such variations were observed all over the year.
During a typical day, the radon concentration tends to increase during the nighttime and decrease during the daytime. This phenomenon comes from the stratification of the atmosphere 
during nighttime, and from the convections due to the sun radiation's heat during daytime.

\subsection{Radon concentration in the Super-Kamiokande area}

\begin{figure}
	\centering
	
	%\begin{subfigure}{\linewidth}
	%	\centering
	%	\includegraphics[width=\linewidth]{Detect20}
	%	\caption{Corridor between the dome and tunnel (detector \#5)} \label{fig:sfig20}
	%\end{subfigure}%

	%\begin{subfigure}{\linewidth}
	%	\centering
	%	\includegraphics[width=\linewidth]{Detect26}
	%	\caption{``Tunnel'' side of the SK dome (detector \#6)} \label{fig:sfig26}
	%\end{subfigure}%
	\includegraphics[width=.9\linewidth]{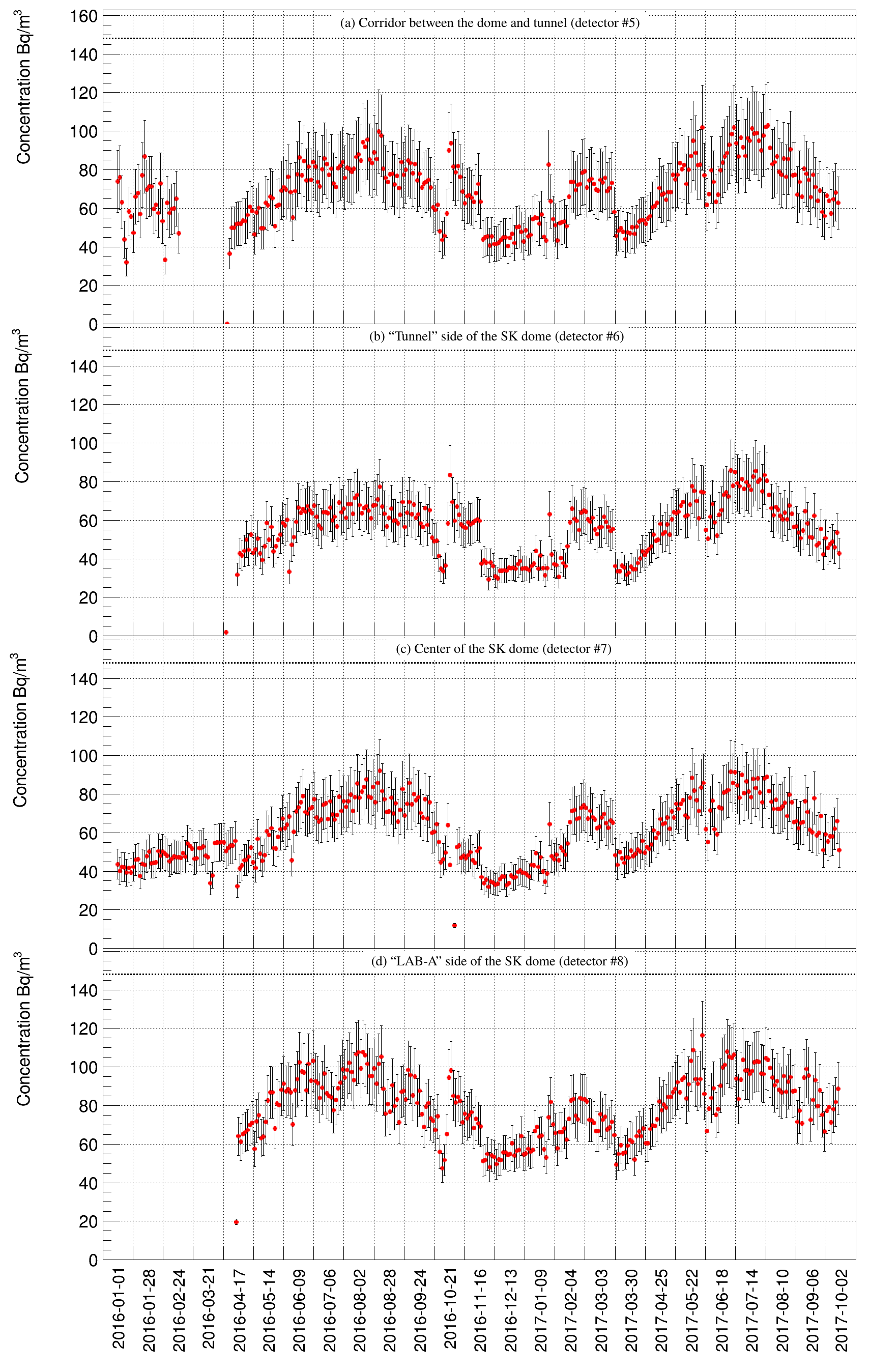}

	\caption{2-days-averaged radon concentration measured in the Super-K dome.
	The dotted line indicates a 148-Bq/m$^3$ U.S. limit for indoor radon concentration \cite{RadonRisk}. 
	The error bars include statistical uncertainties, as well as the uncertainty on the detector's correction factor.
	The detectors \#5, \#6, and \#8 were deployed in April 2016, hence the absence of data before this month.
	\label{fig:DomeConcentrations}}
\end{figure}

As shown in Fig. \ref{fig:KamiokaMap}, several radon detectors are settled in the Super-Kamiokande area.
Fig. \ref{fig:DomeConcentrations} shows the radon concentration measured in the 
dome of the Super-Kamiokande experiment from January 2016 to October 2017. The positions of the detectors in the Super-Kamiokande area are indicated 
in Fig. \ref{fig:KamiokaZoom}. The humidity in this area ranges from 6 to 8 $g/m^{3}$. The periods without data points are periods at which the detectors
were not running due to power cuts or maintenance or because the detector was yet to be deployed. 
The concentration plots show the same patterns of radon-concentration variation, indicating that they are in the same atmosphere. The differences 
in the absolute value of the radon concentration can be explained by the positions of these detectors. In this dome, from the entrance 
of the Super-Kamiokande area to the opposite side, the concentrations are shown by the figures (a), (b), (c), and (d) (detectors \#5, \#6, \#7, and \#8, respectively).
The radon concentration measured near the tunnel-side of the dome is lower than the one measured on the other side of the dome. This phenomenon is due 
to the position of the radon-less air output, which is located on this side of the dome.
The radon concentration in the air inside the tank of Super-Kamiokande is not affected by the variation of radon concentration in the dome \cite{Nakano}.

%\begin{figure}	
%	\centering
	
%	\begin{subfigure}{\linewidth}
%		\centering
%		\includegraphics[width=\linewidth]{Detect15}
%		\caption{Center of the SK dome (detector \#7)} \label{fig:sfig15}
%	\end{subfigure}%
	
%	\begin{subfigure}{\linewidth}
%		\centering
%		\includegraphics[width=\linewidth]{Detect27}
%		\caption{``LAB-A'' side of the SK dome (detector \#8)} \label{fig:sfig27}
%	\end{subfigure}%

%	\caption{2-days-averaged radon concentration measured in the Super-K dome.
%	The dotted line indicates a 148-Bq/m$^3$ U.S. limit for indoor radon concentration \cite{RadonRisk}. 
%	The detector used in Fig. \ref{fig:DomeConcentrations}(\subref{fig:sfig27})
%	was deployed in April, hence the absence of data before this month.
%	\label{fig:DomeConcentrations2}}
%\end{figure}

\begin{figure}
	\centering
	
	\begin{subfigure}{\linewidth}
		\centering
		\includegraphics[width=\linewidth]{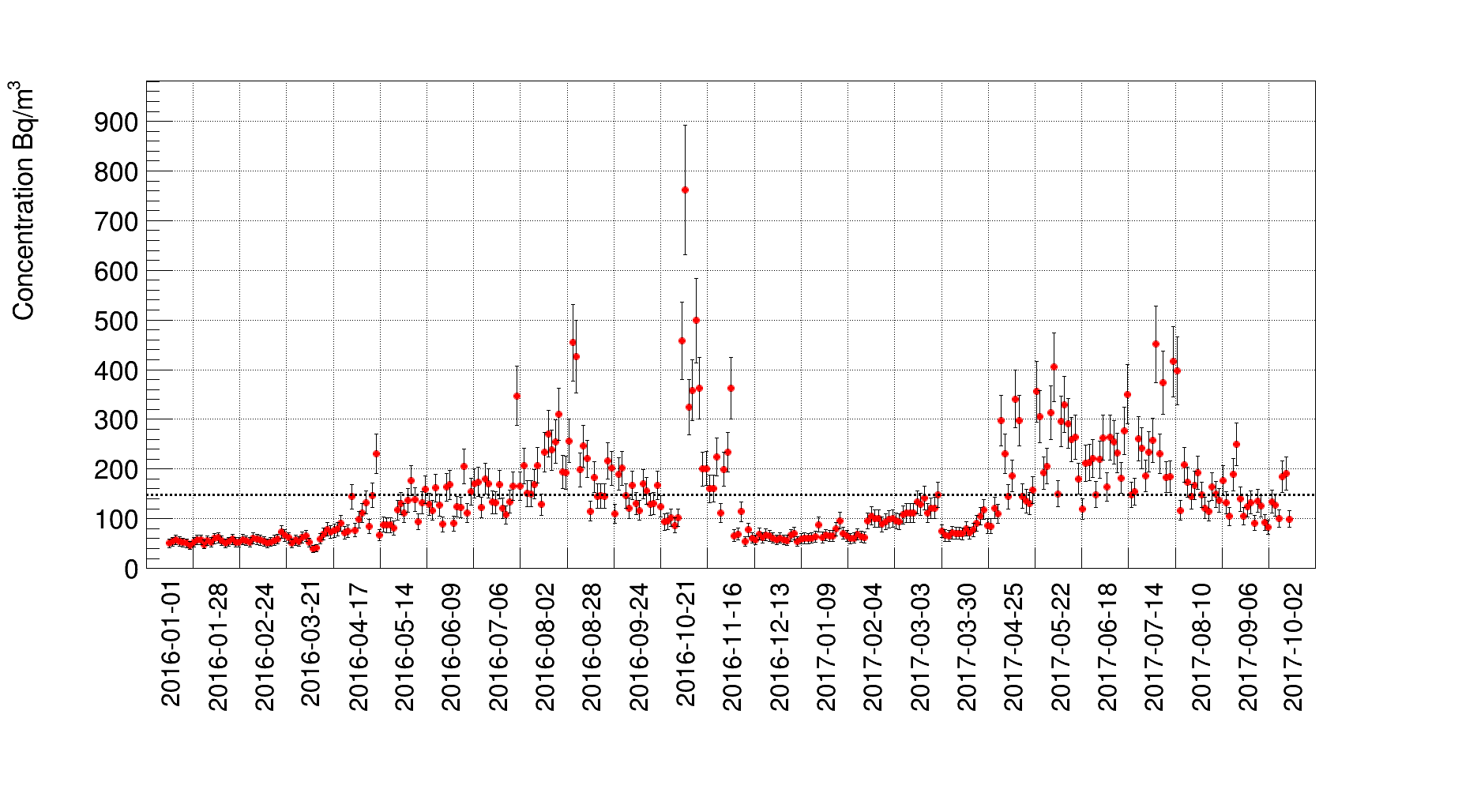}
		\caption{2-days-averaged from January 2016 to October 2017} \label{fig:sfig17}
	\end{subfigure}%
	
	\begin{subfigure}{\linewidth}
		\centering
		\includegraphics[width=\linewidth]{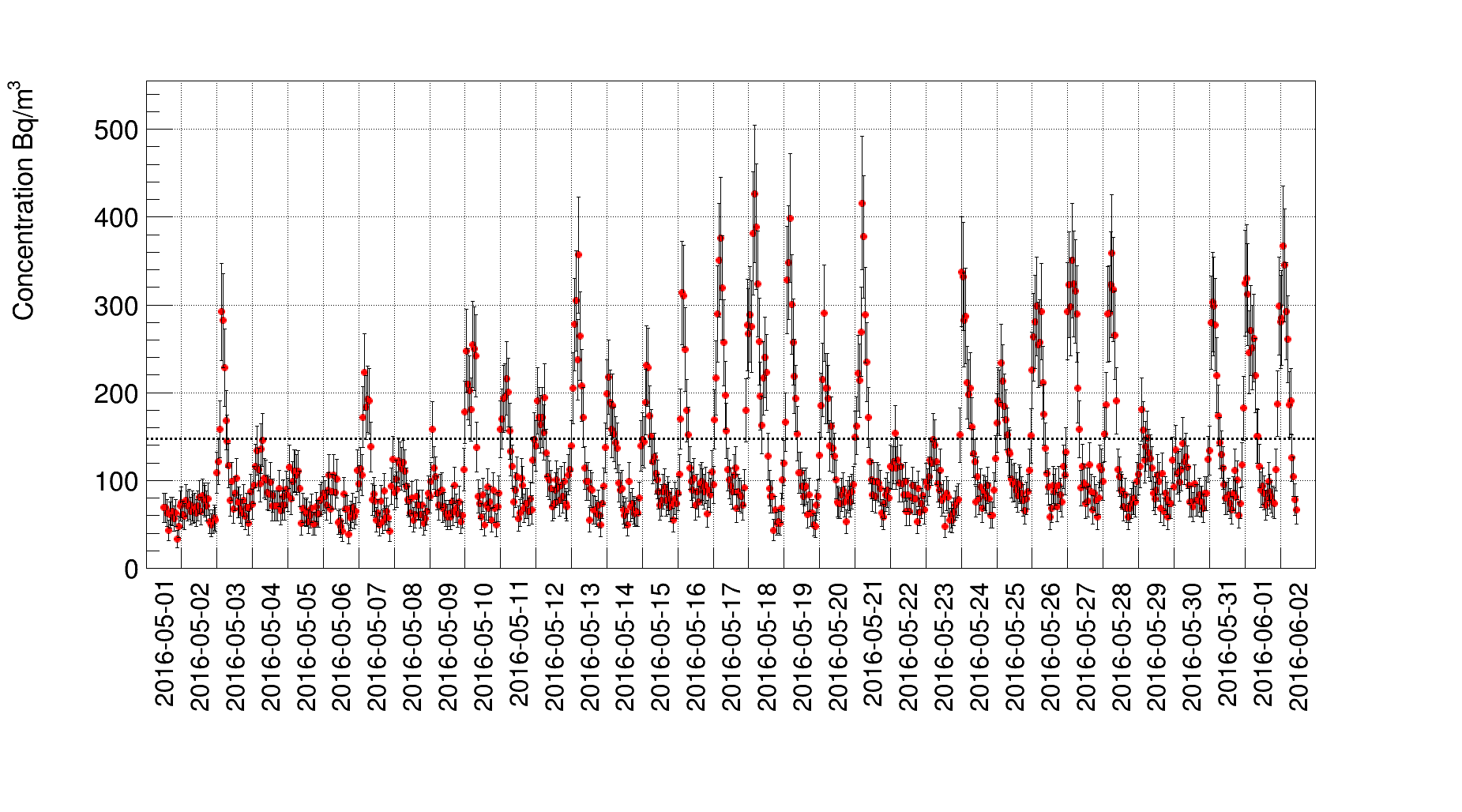}
		\caption{Hourly-averaged over May 2016} \label{fig:sfig17h}
	\end{subfigure}%

	\caption{Radon concentration measured in the entrance room of the Super-K dome (detector \#9).
	The dotted line indicates the 148-Bq/m$^3$ U.S. limit for indoor radon concentration \cite{RadonRisk}.
	The error bars include statistical uncertainties, as well as 
	the uncertainty on the detector's correction factor.
	\label{fig:SKTunnelConcentrations}}
\end{figure}

A detector is settled in the entrance room of the Super-Kamiokande dome (detector \#9). This room is closed by two iron doors, and acts as a transfer airlock
between the tunnel and the dome, in order to prevent the radon-rich air from the tunnel to enter in the dome. Fig. \ref{fig:SKTunnelConcentrations}(\subref{fig:sfig17}) 
shows the 2-days-averaged concentration measured by this detector from January 2016 to October 2017. Fig. \ref{fig:SKTunnelConcentrations}(\subref{fig:sfig17h}) 
shows the hourly-averaged concentration measured by the detector over May 2016. 

Daily fluctuation of the radon concentration is observed. The radon concentration increases during daytime, and decreases during nighttime. This phenomenon 
is due to the radon-rich air from the tunnel entering in the room when workers (including shifters) enters in the Super-Kamiokande area. During nighttime the radon concentration decrease 
as the injection of radon-rich air from the tunnel is reduced. A confirmation of this interpretation arises from the observation of the low radon concentration during Japanese
national holidays (like on May 3rd, May 4th, and May 5th in Fig. \ref{fig:SKTunnelConcentrations}(\subref{fig:sfig17h})), and during Saturday and Sunday, when few 
workers enter in the area.

On October 20th 2016, a sudden increase of the radon concentration can be observed in Fig. \ref{fig:SKTunnelConcentrations}(\subref{fig:sfig17}), as well as in 
Fig. \ref{fig:DomeConcentrations}. The door between Atotsu tunnel and the entrance room was left open overnight while the radon concentration in the tunnel was about $1700$ $Bq/m^{3}$, 
as shown in Fig. \ref{fig:TunnelConcentrations}(a). In Fig. \ref{fig:DomeConcentrations} and Fig. \ref{fig:SKTunnelConcentrations}(\subref{fig:sfig17}), another radon concentration increase is observed over February 2017, due to an interruption of the 
aeration system of the entrance room.

\section{Summary}

\begin{table}[hbt!]
	\small
	\centering
	\begin{tabular}{ | c   c | c | c | c | c | }  
		\hline
		\multicolumn{2}{| c | }{Area}				& Super-K		& Super-K		& Atotsu 		& Fresh air 	 	\\
		\multicolumn{2}{| c | }{ }				& Dome			& Entrance		& Tunnel		& pipe 		 	\\
		\hline 
		\multicolumn{2}{| c | }{Detectors}			& 4			& 1			& 1			& 2		 	\\
		\hline
		Averaged  	& Year					& $64.2 \pm 0.3$	& $137.7 \pm 1.2$	& $1082.6 \pm 10.5$	& $28.6 \pm 0.3$ 	\\
		\cline{2-6}
		radon		& Winter				& $56.0 \pm 0.6$ 	& $72.1  \pm 1.3$ 	& $209.6  \pm 5.0$ 	& $14.1 \pm 0.3$ 	\\
		\cline{2-6} 
		conc.		& Spring				& $63.8 \pm 0.6$ 	& $146.9 \pm 2.6$ 	& $1261.1 \pm 23.3$	& $21.5 \pm 0.5$ 	\\
		\cline{2-6} 
		(Bq/m$^3$)	& Summer				& $79.2 \pm 0.8$ 	& $194.4 \pm 3.5$	& $1739.1 \pm 32.1$	& $50.7 \pm 1.0$	\\
		\cline{2-6} 
				& Fall					& $55.2 \pm 0.5$  	& $136.0 \pm 2.5$	& $791.0  \pm 14.3$ 	& $22.7 \pm 0.4$	\\
		\hline 

	\end{tabular}

	\caption{Summary table of the averaged radon concentrations in different important areas of the mine. 
	The second row indicates the number of detectors measuring the atmosphere of each area. The third row indicates the yearly-averaged 
	radon concentration. The fourth to seventh rows indicate the four seasonly-averaged radon concentrations. Winter is defined as between December 21st and March 21st;
	Spring as between March 21st and June 21st; Summer as between June 21st and September 21st; and Fall as between September 21st and December 21st.
	\label{tab:SummaryTable}}
\end{table}

New 1-L radon detectors and Raspberry Pi electronics have been developed for radon monitoring at the Kamioka Observatory. As of December 2017, twenty one 1-L detectors using Raspberry Pi 
electronics are used to monitor the radon concentration in the air in the mine. A total of 28 of these devices are used within the research facility (the Kamioka mine and surface 
buildings). 
These detectors and electronics have now been used for more than one year with stable data acquisition, as summarized in Tab. \ref{tab:SummaryTable}. 
We observed seasonal variations of the radon concentration in the air of the mine. Day/night variations
of the radon concentration in the fresh air brought into the mine have also been observed. 

\section*{Acknowledgment}

The authors would like to thank Mitsuhiro Nakamura, from the Fundamental Particle Physics Laboratory, Graduate School of Science of Nagoya University, for his help in the development of the pre-amplifier circuit.  
We would like to express our sincere gratitude to Prof. Takao Iida, from the Graduate School of Engineering of Nagoya University who gave us adequate advice and discussion on the development of the radon detector. 
We also would like to thank the Super-Kamiokande collaboration for their help in conducting this study. We gratefully acknowledge the cooperation of the Kamioka Mining and Smelting Company.

\end{document}